%% file: paper.tex
\documentclass[11pt,a4paper]{article}

\input{settings_2.tex}   
\input{shortcuts.tex}          
\graphicspath{{./figures/}}                

\begin{document}

\title{
Calomplification --- The Power of Generative Calorimeter Models
}

\author[a]{S.~Bieringer}
\author[b]{A.~Butter}
\author[a]{S.~Diefenbacher}
\author[c]{E.~Eren}
\author[c]{F.~Gaede}
\author[a]{D.~Hundhausen}
\author[a]{G.~Kasieczka}
\author[d,e]{B.~Nachman}
\author[b]{T.~Plehn}
\author[f]{M.~Trabs}

\affiliation[a]{Institut f\"ur Experimentalphysik, Universit\"at Hamburg, Luruper Chaussee 149, 22761 Hamburg, Germany}
\affiliation[b]{Institut f\"ur Theoretische Physik, Universit\"at Heidelberg, Philosophenweg 16, 69120 Heidelberg, Germany}
\affiliation[c]{Deutsches Elektronen-Synchrotron DESY, Notkestr. 85, 22607 Hamburg, Germany}
\affiliation[d]{Physics Division, Lawrence Berkeley National Laboratory, Berkeley, CA 94720, USA}
\affiliation[e]{Berkeley Institute for Data Science, University of California, Berkeley, CA 94720, USA}
\affiliation[f]{Department of Mathematics, Karlsruhe Institute of Technology, Englerstr. 2, 76131 Karlsruhe, Germany}

\emailAdd{sebastian.guido.bieringer@uni-hamburg.de}

\arxivnumber{2202.07352}

\abstract{Motivated by the high computational costs of classical simulations, machine-learned generative models can be extremely useful in particle physics and elsewhere. They become especially attractive when surrogate models can efficiently learn the underlying distribution, such that a generated sample outperforms a training sample of limited size. This kind of GANplification has been observed for simple Gaussian models. We show the same effect for a physics simulation, specifically photon showers in an 
electromagnetic calorimeter.}


\maketitle

\section{Introduction}
\label{sec:intro}

Particle physics research at colliders is defined by extremely large datasets combined with precision simulations, from first principles all the way to a detailed detector simulation. A reliable generation and simulation chain is crucial to link measurements to fundamental properties of elementary particles. This chain is factorized into two main parts, event generation based on a fundamental Lagrangian and perturbative or non-perturbative quantum field theory, and detector simulations describing the interactions of relativistic particles with the detector. For the upcoming runs of the Large Hadron Collider (LHC), both parts need to gain significantly in speed, to keep up with the size of experimental datasets. One way to achieve this speed gain is to apply modern machine learning (ML) to all levels of the simulation chain. A key tool in this speed-improvement program is deep generative neural networks (NNs) that learn to emulate slower physics-based simulations, replacing the underlying physics by fast and accurate \emph{surrogate models}.

A foundational question with NN surrogate models is, what are the advantages of using the fast simulation compared with the original dataset used for training? Or specifically, how many more events can we sensibly generate from these models before we are limited, for instance, by the training statistics?  Without any additional information, we would expect that the statistical power of a generated dataset is at most the same as the dataset used for training. A larger generated sample than the training dataset will then include successively less information per event than the training data, and eventually the information in the generated events will saturate and be dominated by 
limitations from the network architecture and training. With this pattern in mind~\cite{Butter:2020qhk}, we can define an amplification or GANplification factor~\cite{hao2019data2,2022arXiv220104315A} in terms of an effective sample size for a given surrogate model. 

GANplification arises, intuitively, from the fact that neural networks work like classical parametric fits~\cite{Bellagente:2021yyh,Chahrour:2021eiv}, and they are particularly effective when we want to interpolate in many dimensions. This feature is behind the success of the NNPDF parton densities~\cite{DelDebbio:2004xtd} as the first mainstream ML-application in particle theory.

Formally, this fit-like effect is one source of inductive bias, where the underlying assumption is that physics probability densities are smooth. Especially in particle physics, it should be possible to employ other inductive biases, such as symmetries or fundamental invariances in datasets~\cite{Krippendorf:2020gny,Barenboim:2021vzh,Dillon:2021gag,Lester:2021kur,Tombs:2021wae,2112.05722}. Fast detector simulations benefit from the fact that we can factorize the problem into pieces. Surrogate models are trained to produce a detector response for each outgoing particle.  For example, if there is an event with $M$ outgoing particles, each one will be attached to a sampling from the surrogate model.  If the training set has $N$ detector interactions, additional combinatorial factors appear for choosing $N$ out of $M$ different events that could be created. These factors can lead to another statistical amplification. Finally, surrogate models with valid inductive biases require far fewer parameters to specify than the original dataset, so there will also be a benefit in the required disk space. 

The goal of this paper is to study the statistical amplification of deep generative models, focusing on interpolation from the smoothness inductive bias, for detector simulations as a realistic and highly relevant application. Fast surrogate models for detector simulations have been developed~\cite{by_example,calogan1,calogan2, Vallecorsa:2019, Carrazza:2019cnt, gan_phasespace,DijetGAN2,Chekalina:2018hxi, fast_accurate, Deja:2019, deOliveira:2017rwa, monkshower,Howard:2021pos} and improved~\cite{aachen_wgan1, aachen_wgan2,Backes:2020vka, Belayneh:2019vyx,Buhmann:2020pmy,Buhmann:2021lxj,Buhmann:2021caf,Krause:2021ilc,Krause:2021wez,Khattak:2021ndw, Kansal:2021cqp,Hariri:2021clz,Rehm:2021zow,Rehm:2021zoz, Rehm:2021qwm} to the level that they are ready to be used in the upcoming LHC runs. In fact, the ATLAS Collaboration has already integrated a Generative Adversarial Network (GAN) into its fast calorimeter simulation and will use it to generate over a billion events~\cite{ATLASShowerGAN,ATLAS:2021pzo}. Initial studies exist on quantifying uncertainties of generative models in event generation~\cite{Butter:2021csz}, but there has not yet been a study of the fundamental benefits of deep generative surrogates applied to detector simulations.

In this paper, we study statistical amplification in the context of photon showers in an 
electromagnetic calorimeter for a GAN-like generative model (Calomplification). However, the method can be applied to gauge the merit of generative surrogates whenever the underlying distribution can be accessed either through a large number of samples or analytically. We expect similar results in all cases where the smoothness assumption on the underlying density distribution is valid. 

The paper is organized as follows.  In section~\ref{sec:data}, we start by introducing our data set and the established generative Variational Autoencoder-GAN (VAE-GAN) architecture adapted to this simulation~\cite{Buhmann:2020pmy}. Next, we describe our treatment of the comparison between generated and truth samples and the relevant observables in section~\ref{sec:metrics}. We then present the amplification effects of the generative networks in section~\ref{sec:results}. This comparison includes an estimate of the effective sample size to the information encoded and a comparison to standard density estimators. In section~\ref{sec:conclusions}, we briefly summarize our promising findings.
   
\section{Dataset and model}
\label{sec:data}
    
\begin{figure}[t]
    \includegraphics[width=0.226\linewidth]{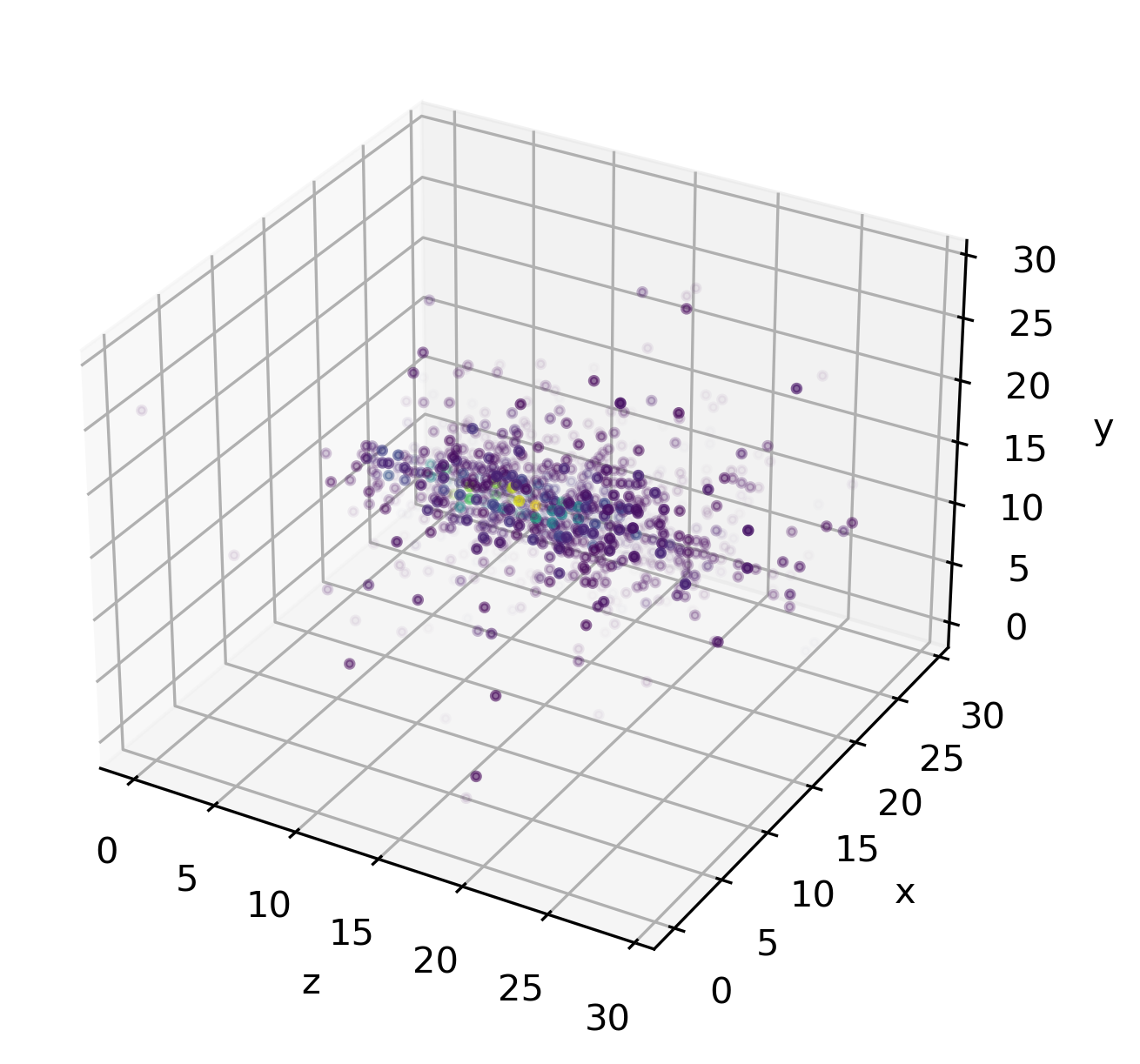}
    \includegraphics[width=0.229\linewidth]{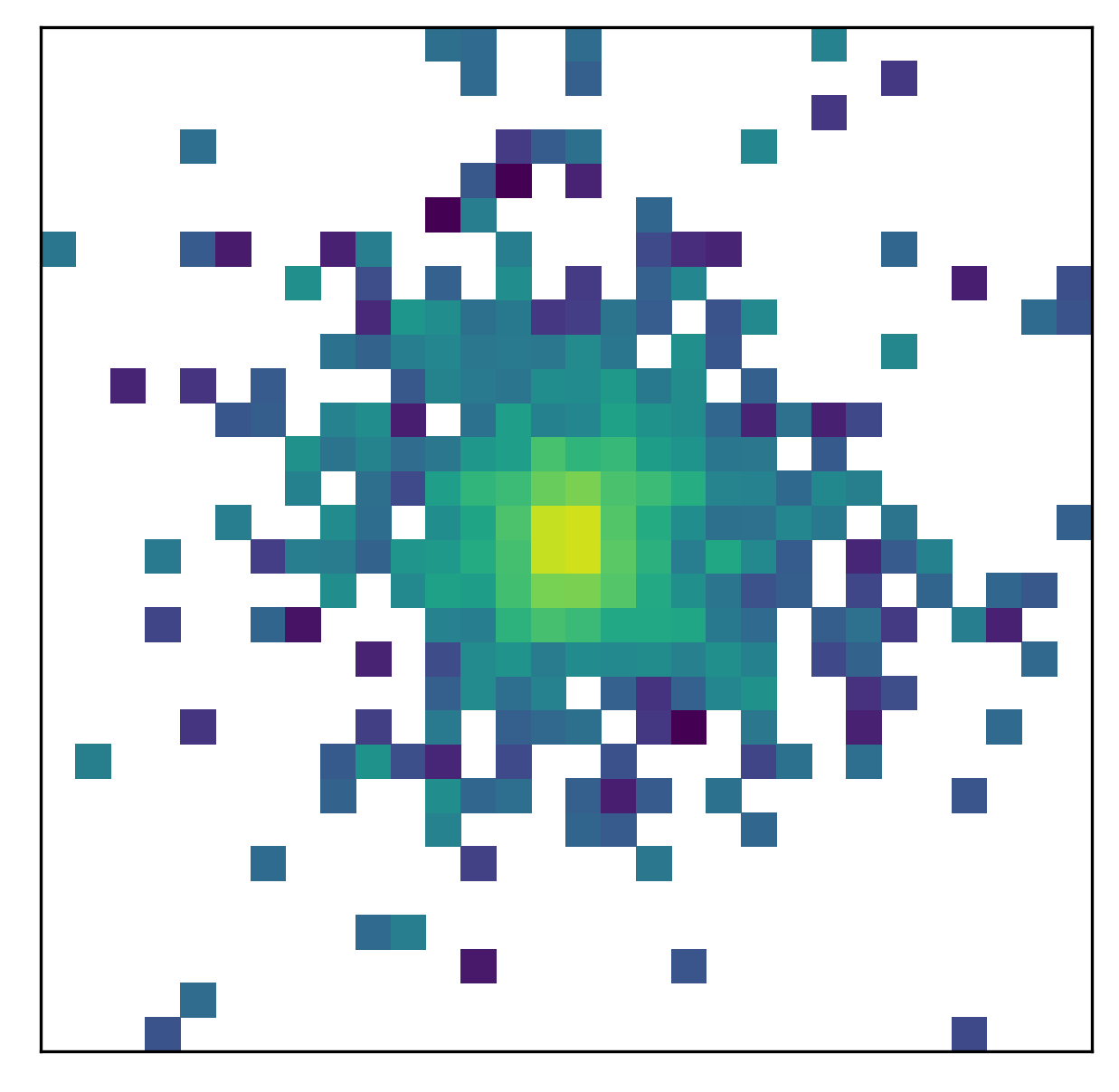}
    \includegraphics[width=0.225\linewidth]{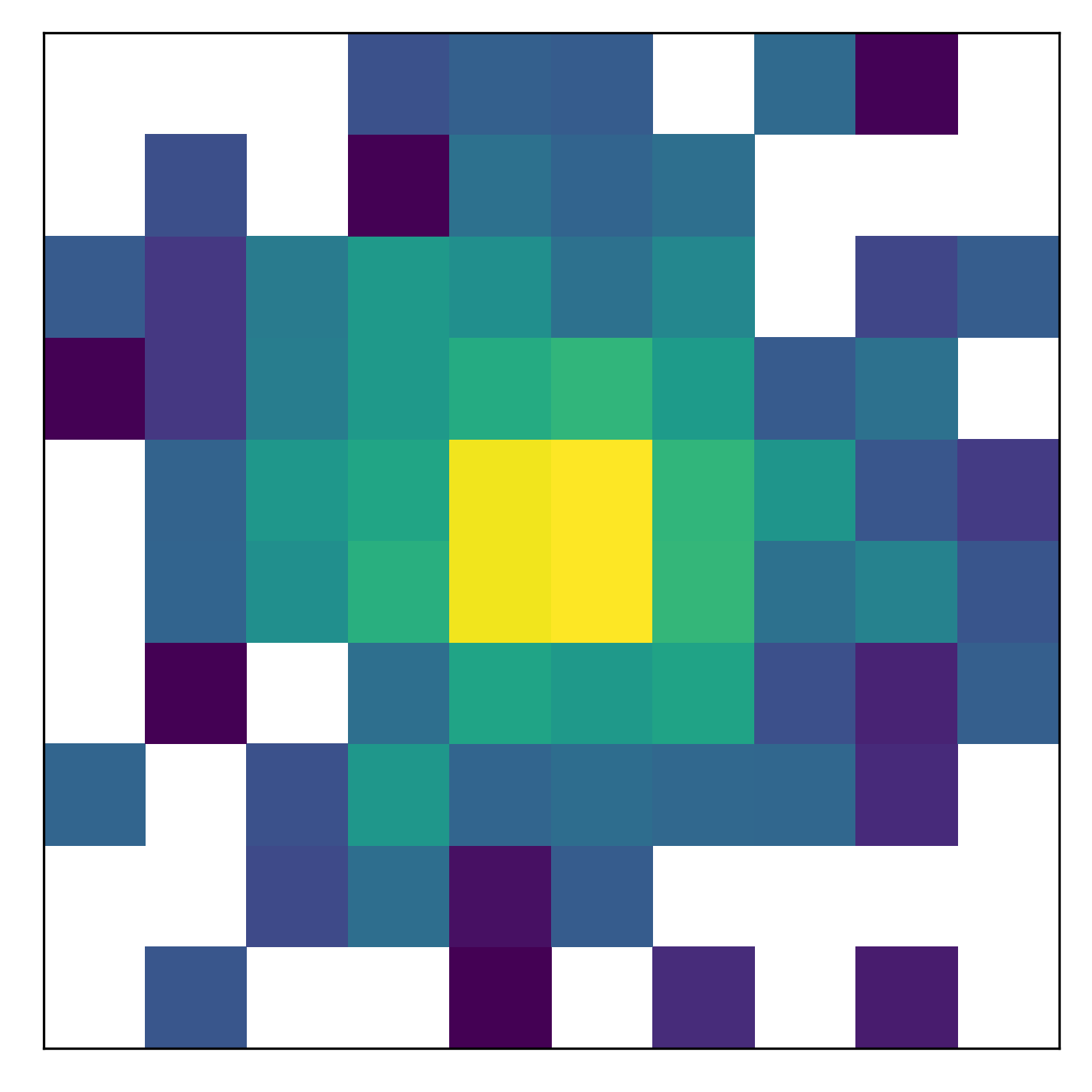}
    \includegraphics[width=0.289\linewidth]{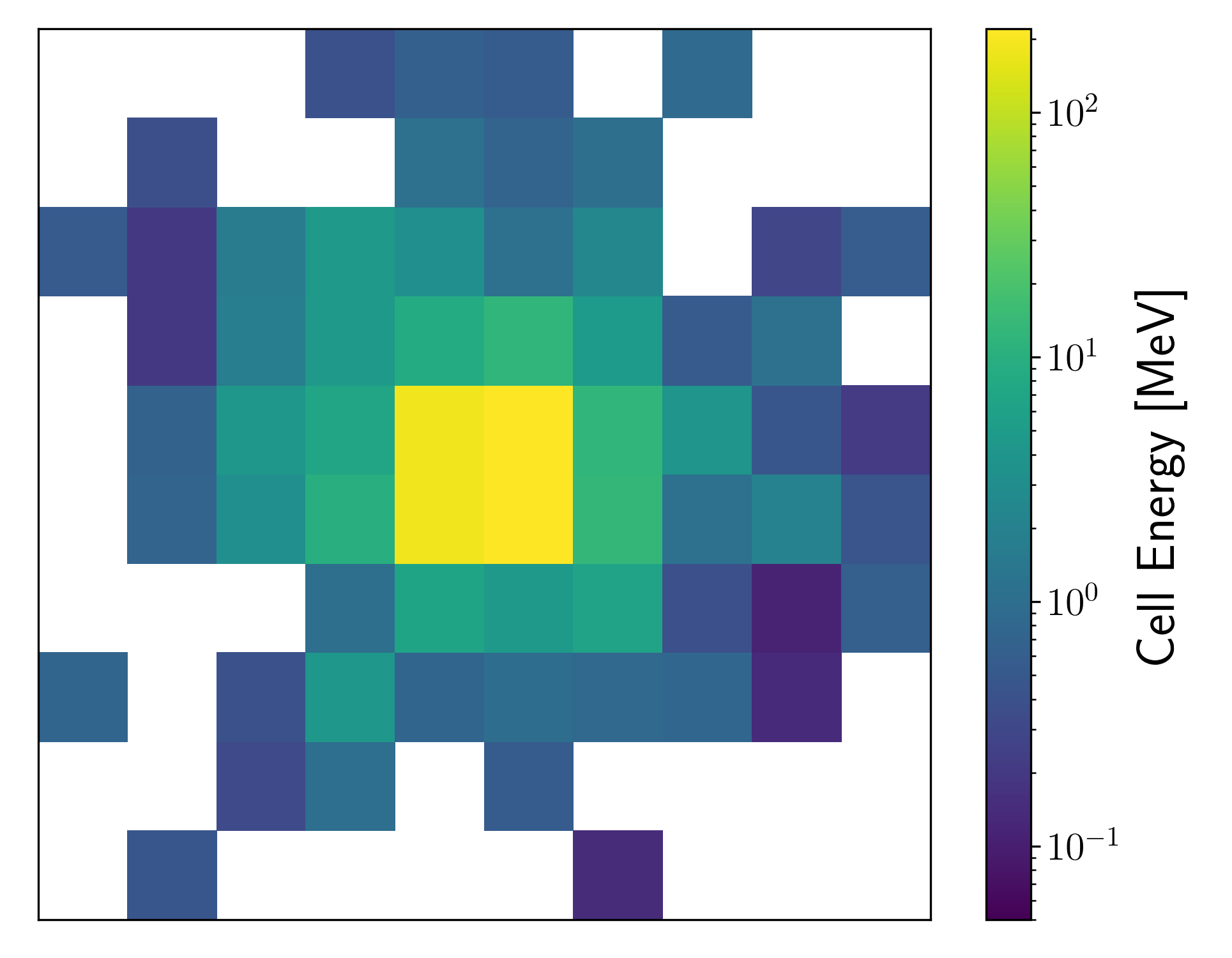}
    \caption{Illustrated transformation of the original calorimeter images from left to right. All histograms feature a logarithmic color coding, with an equal scaling for the $10\times10$ images. The final step of cutting below half the MIP energy is applied for evaluation only.}
    \label{fig:dataset}
\end{figure}

\begin{figure}[b!]
    \centering
    \includegraphics[width=0.85\linewidth]{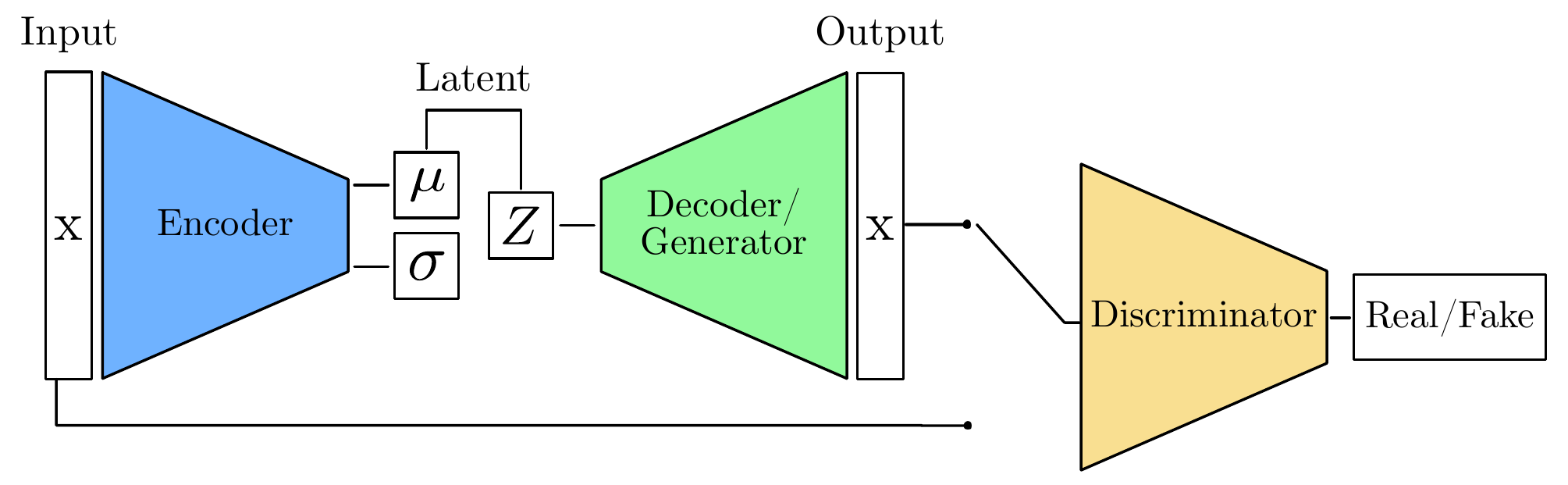}
    \caption{Illustration of the VAE-GAN architecture. The encoder and decoder form a VAE setup, while the decoder can also be understood as a GAN generator. The discriminator acts as a binary classifier, as in a classical GAN.}
    \label{fig:vaegan}
\end{figure}

The International Large Detector (ILD)~\cite{ILD-IDR} is one of two detector concepts proposed for the International Linear Collider (ILC).  It is optimized towards the Particle Flow analysis concept for optimal global event reconstruction~\cite{PFA:2009,Marshall:2015rfa}. It combines high-precision tracking and vertexing capabilities with very good hermiticity and highly-granular electromagnetic and hadronic calorimeters (ECal/HCal). We choose one of its two proposed electromagnetic calorimeters, the Si-W ECal, for our dataset. It consists of 30 active silicon layers in a tungsten absorber stack with 20~layers of 2.1~mm and 10~layers of 4.2~mm thickness. The silicon sensors have a cell size of $5\times 5~\text{mm}^2$. 
    
ILD uses iLCSoft~\cite{ilcsoft} for detector simulation, reconstruction, and analysis. The \geant~\cite{ALLISON2016186} simulation uses a realistic detector model implemented in DD4hep~\cite{dd4hep}. Photons are shot into the ECal barrel at a perpendicular incident angle. We project the cells with energy depositions (hits) onto a rectangular grid of  $30\times30\times30$~cells. We choose photon showers, because their structure is more regular and faster to learn than the structure of pion showers~\cite{Buhmann:2021caf}. 

To develop a high-precision generative model, we fully simulate 268k photon showers with a fixed energy of 50~GeV. From the full set, 50k showers are randomly selected for training (1k) and for the evaluation of the network performance (all 50k). Whenever we need to estimate the generative model uncertainty, we train the network on five sets of 1k training samples. To include an error estimate for the evaluation samples, we use five sets of 5k or 10k evaluation showers, chosen as subsets of these 50k showers. The remaining 218k showers are used as a high-statistics estimate of the truth distribution. 

To simplify the training of our precision-generative model, we reduce the dimensionality of the images to $10\times10$ pixels, summing along the beam axis and pooling $3\times3$ patches of the resulting 2D-images. The process is illustrated in figure~\ref{fig:dataset}. We will always refer to the combined calorimeter cells as pixels of the calorimeter image. The reduction allows us to obtain a powerful model from a small training set, such that the majority of the data can be used to estimate the truth distribution. Finally, we apply a cut at 0.1~MeV, which corresponds to the most probable energy deposition of a minimal ionizing particle (MIP). Cell energies below have a low signal to noise ratio. To aid the network training, this cut is not present in the training data, but applied on the full set of generated and reference data.

In comparison to studies done in context of proposed, high-granularity tracking calorimeters, these simplifications seem extensive. However, for simulation of the current ATLAS detector the AtlFast3 simulation tool~\cite{ATLAS:2021pzo} uses $300$ individual GANs, each generating only one $eta$-slice of the calorimeter. Each network generates a 2D-image in the radius-$phi$-plane of the detector and only $1000$ events are generated to learn the highest energy samples. Albeit, the models are trained including ten-thousands of lower energy samples. We see that in order to facilitate a comparison to a large validation set, the task has not been simplified further than in current applications. \medskip

Our generative architecture is a VAE-GAN~\cite{vae_gan_2016}, closely related to the network developed for precision simulations of photon showers~\cite{Buhmann:2020pmy} and illustrated in figure~\ref{fig:vaegan}. It closely resembles a standard VAE setup, but deviates in its use of a GAN-like discriminator as a substitute for the usual element-wise reconstruction loss. The loss function is
\begin{align}
\loss_\text{VAE-GAN} &= \loss_\text{GAN}
+ \underbrace{D_\text{KL}\left( q_\text{encoder}(z|x) | p(z) \right)}_{\mathcal{L}_\text{prior}} \notag \\
\text{with} \qquad 
\loss_\text{GAN} &= \mathbb{E}_{x\sim p_\text{data}(x)} \big[ \log D(x) \big] + \mathbb{E}_{z\sim q_\text{encoder}(z|x)} \big[ \log (1-D(G(z))) \big] \; ,
\label{eq:loss}
\end{align}
where $p(z) = \mathcal{N}(0,I)$. We maximize $\loss_\text{GAN}$ during discriminator optimization. Every two discriminator steps, we update generator and encoder by minimizing the full loss function, $\loss_\text{VAE-GAN}$, \ie the generator dependent part of $\loss_\text{GAN}$ and the second term $\mathcal{L}_\text{prior}$. The prior loss regularizes the latent space and allows us to sample from $p(z) = \mathcal{N}(0,I)$ during generation. For generator updates, we recast the GAN loss to  $-\log D(G(z))$ to ensure efficient training for early epochs~\cite{goodfellow}. In every update step we sample $z$ only once per input $x$. Using a GAN-like discriminator is essential, as the range of pixel values covers multiple orders of magnitude. For such images, the element-wise reconstruction loss is dominated by the central, high-energy pixels.

\begin{figure}[t]
    \centering
    \begin{subfigure}[b]{0.9\textwidth}
        \centering
        \resizebox{.7\textheight}{!}{%
        \input{figures/lagan_gen}%
        }
        \caption{Generator}
        \label{fig:generator_setup}
    \end{subfigure}
    \break
    \begin{subfigure}[b]{0.9\textwidth}
        \centering
        \resizebox{.7\textheight}{!}{%
        \input{figures/lagan_disc}%
        }
    \caption{Discriminator}
    \label{fig:discriminator_setup}
    \end{subfigure}
\caption{Generator and discriminator setup including parameter space sizes in between operations. The feed-forward for both networks proceeds left to right.}
\label{fig:lagan}
\end{figure}
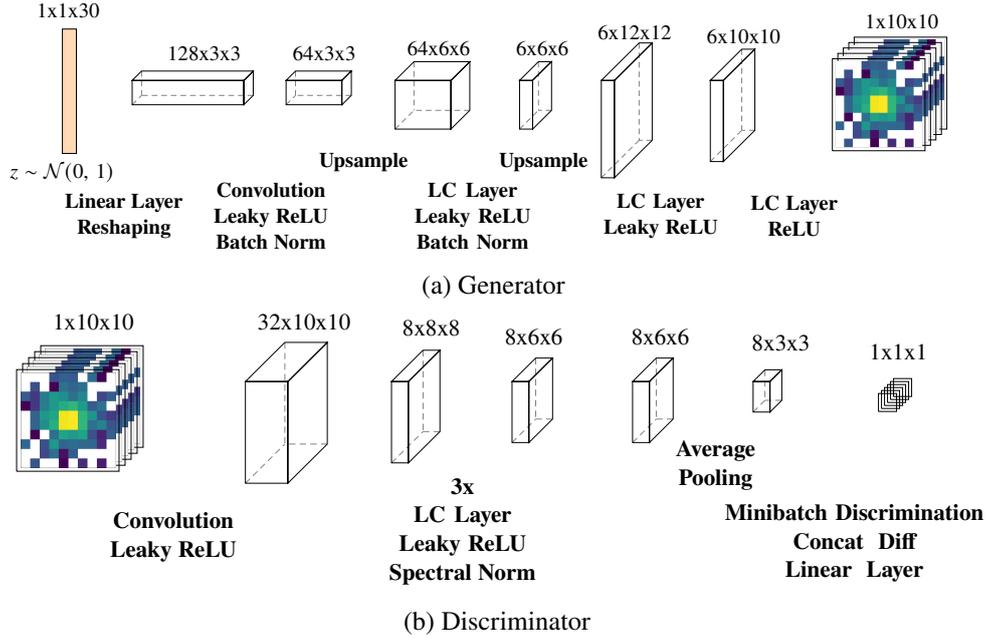

The GAN-like part of our network is modeled after the LAGAN~\cite{by_example} illustrated in figure~\ref{fig:lagan}. Unlike many standard applications of convolutional networks, the LAGAN features locally connected layers. Other than convolutional layers, these have the flexibility to account for the missing translation symmetry in calorimeter images. A few changes are made to the original LAGAN setup, including modifying the dimensionality of our network layers to conform to our image
and latent space sizes. We also replace batch norm by spectral norm in the discriminator~\cite{spec_norm} to further stabilize the training. The discriminator uses the difference between reconstructed images and the corresponding training images as an additional input to the final, fully connected layer. For the training images themselves, this difference vector is zero. We apply label smoothing to prevent vanishing gradients from an overconfident classifier. Supplementing the information gained from the images themselves with locally connected layers and mini-batch discrimination~\cite{Salimans2016} ensures better consistency between training and generated images.

The encoder network uses a convolutional input and two convolutional hidden layers, applying Leaky Rectified Linear Unit activation (LeakyReLU)~\cite{Maas13rectifiernonlinearities} after the first two and ReLU after the third layer. The output of the encoder's convolutional part is fed to two separate linear layers, defining the mean and $\log \text{var}$ values of the Gaussian VAE latent space.
    
Our network is implemented in \pytorch1.8.0~\cite{pytorch} and trained on Nvidia P100s using the \Adam optimizer~\cite{adam} with a learning rate of $8 \times 10^{-6}$ for all networks. Each training on 1k showers is run for 24h, amounting to around 50000 epochs. For epochs after 40000, the distributions of the (i) pixel energy sum (visible energy), (ii) highest pixel energy (peak energy), (iii) per-pixel energy, and, (iv,v) the pixel position weighted by the pixel energy (center of gravity) in a given direction,
\begin{align}
\big\{ \; E_\text{vis}, E_\text{peak}, E_\text{pixel}, \text{CG}_{x,y} \; 
\big\}  \; ,
\label{eq:obs}
\end{align}
are estimated using histograms of 96000 generated images. The histograms feature 100 bins and constant ranges. Finally, we select the epoch with the best agreement between the generated and training distributions averaged over all five observables, in terms of the measure discussed in the next section. This procedure is repeated for three independent trainings per set of training samples, and we draw a VAE-GAN sample in equal proportions from the resulting three models. We are aware that three independently trained models are not statistically sufficient to define a reliable standard deviation, but we have found them to be very helpful and sufficient in estimating the stability of the network training. The results in section~\ref{sec:results} feature the standard deviation on the five different training sets. Whenever we show $E_\text{pixel}$ we apply an additional minimum cut of 5~MeV, as will be discussed in detail in the next section.

\section{Sample comparison}
\label{sec:metrics}

\begin{figure}[t]
    \includegraphics[width=0.24\linewidth]{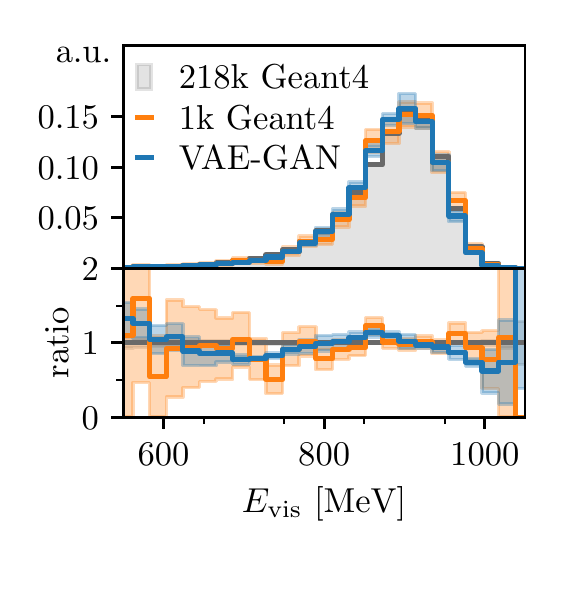}
    \includegraphics[width=0.24\linewidth]{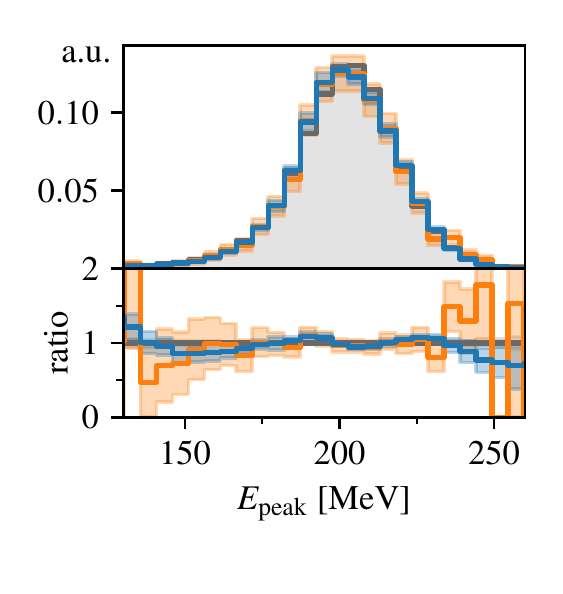}
    \includegraphics[width=0.24\linewidth]{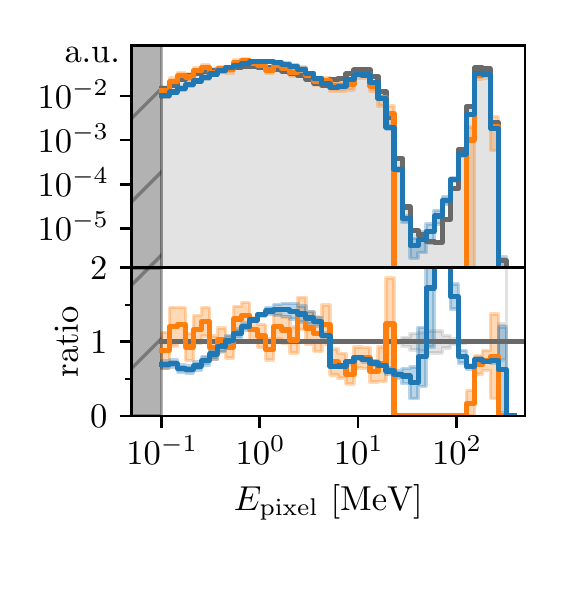}
    \includegraphics[width=0.24\linewidth]{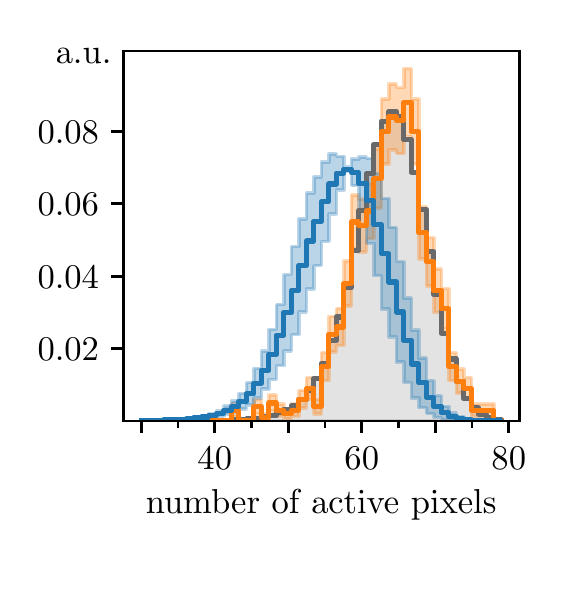}
    \includegraphics[width=0.24\linewidth]{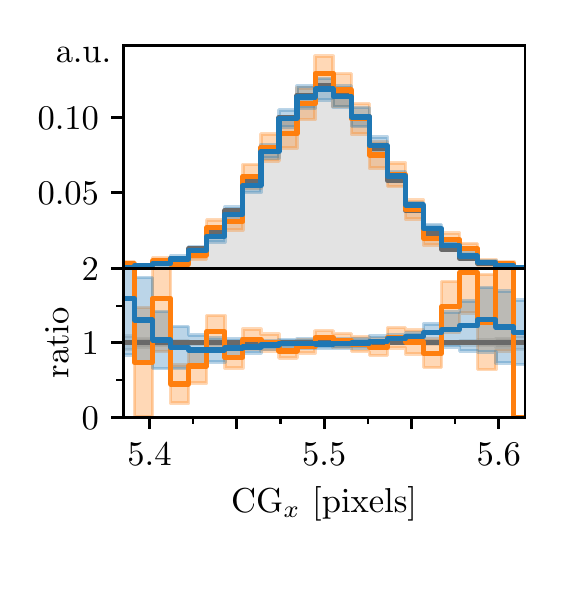} \hfill
    \includegraphics[width=0.24\linewidth]{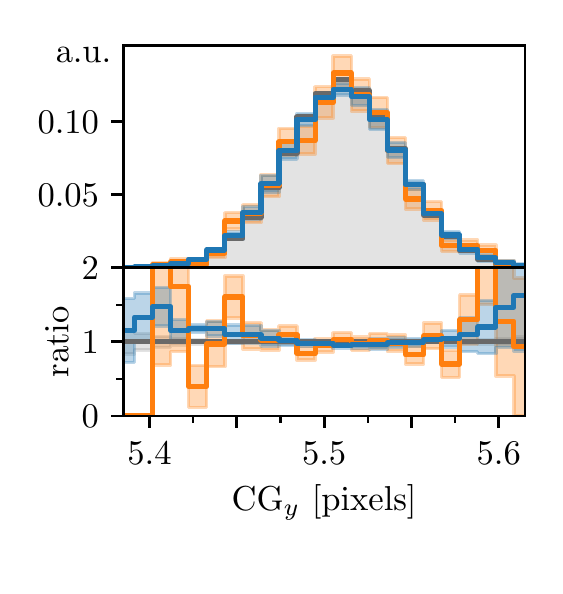} \hfill
    \includegraphics[width=0.24\linewidth]{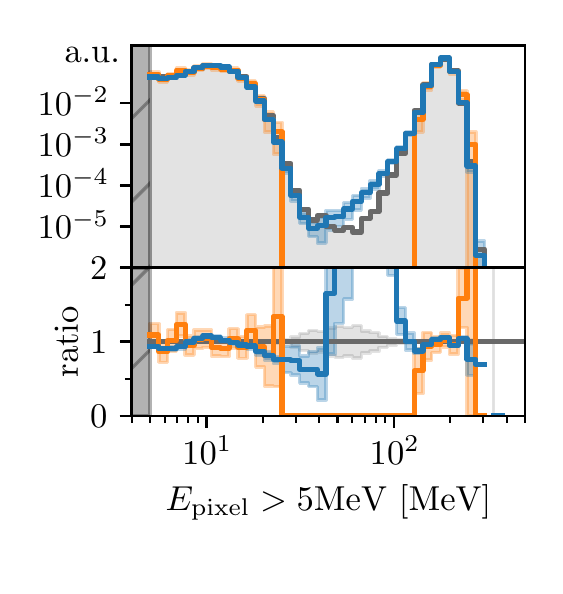} \hfill
    \includegraphics[width=0.24\linewidth]{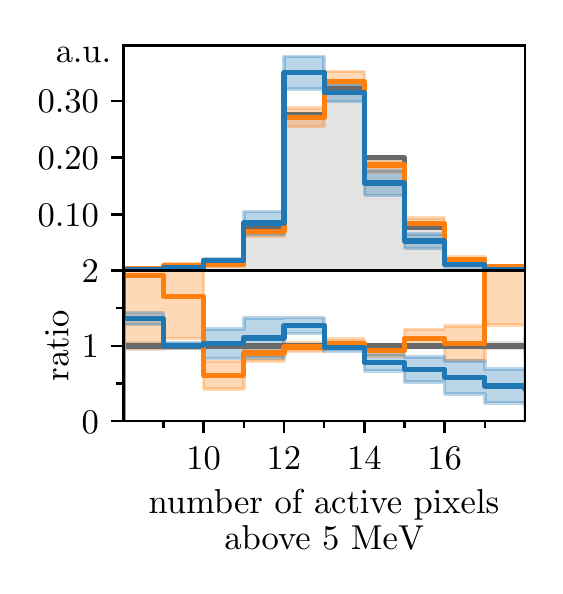}
    \caption{Differential distributions for the observables given in eq.~\eqref{eq:obs} from \geant and from the VAE-GAN-generated images. Errors of the validation set (grey) and the training set (orange) correspond to the Poisson-error per bin, while the uncertainty on the VAE-GAN line (blue) is illustrated by the standard deviation of three independent trainings on the 1k training data. All histograms are normalized, such that all bins add up to one. The insets show the ratio to the high-statistics estimate of the truth distribution.}
    \label{fig:perf}
\end{figure}

To determine the performance of the trained model, we again use distributions of the same five high-level observables as for the training. We compare showers generated by \geant and our VAE-GAN, but now using the high-statistics validation set. Figure~\ref{fig:perf} shows a set of distributions for 1k shower images used for a single VAE-GAN training and 1000k showers from the corresponding generative network. They are compared to the validation set of 218k \geant showers. In addition to the continuous distributions we also show the number of active pixels per image. First, we see that statistical fluctuations of the training set propagate into under- and over-densities of the learned distributions. One prominent difference is the number of active pixels, which can be attributed to the under-estimation of the number of low energy hits below 5~MeV. The remaining learned distributions are smoother and show fewer fluctuations than the training data. For the visible per-pixel energy, the VAE-GAN interpolates into the sparsely populated interval between around 2 and 120~MeV even though the training set does not include a single pixel in this range. Previous work has shown~\cite{Buhmann:2020pmy} how to correct the low-energy behavior through an additional, consecutively trained post-processing network, using an maximum mean discrepancy loss~\cite{NIPS2006_e9fb2eda,gan_phasespace} on the pixel energy spectrum. Here we skip this post-processing and instead focus on the statistical properties of the generated data for visible pixel energies above 5~MeV.
    
\subsubsection*{Quantiles}

\begin{figure}[t]
    \centering
    \includegraphics[width=1\linewidth]{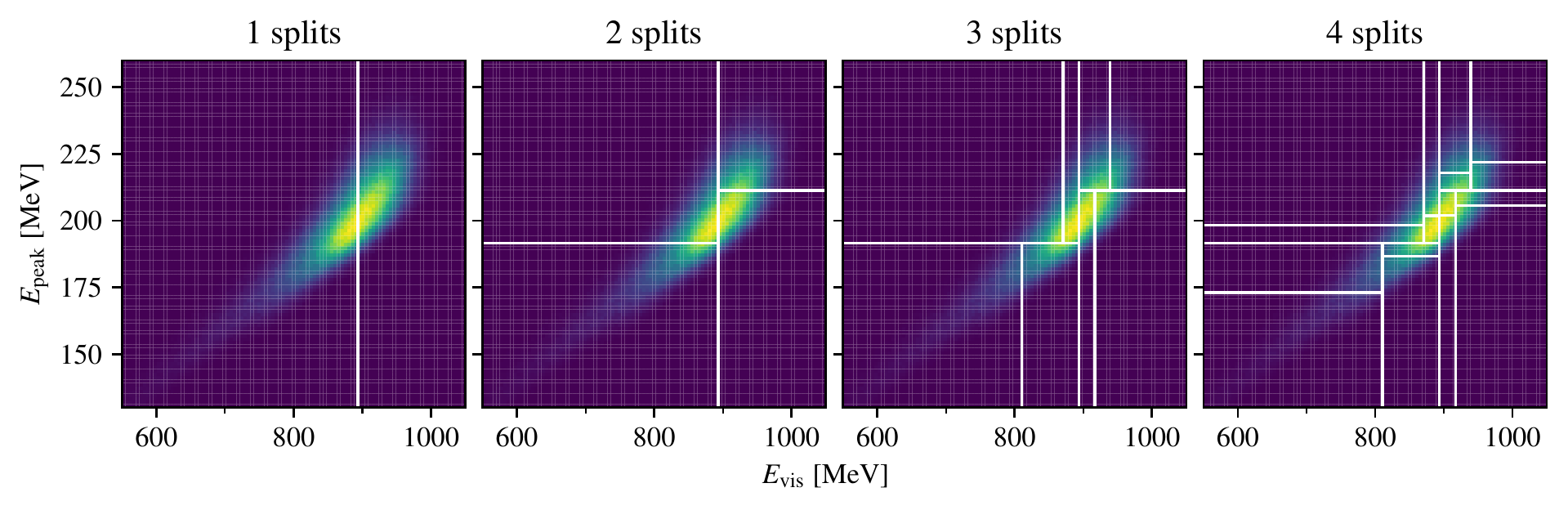}
    \caption{Quantiles developing by splitting of the validation set into subsets of equal size regarding their energy sum \& peak energy.}
    \label{fig:quantiles}
\end{figure}

We now turn to quantifying the efficacy of the VAE-GAN, given the strong performance shown in figure~\ref{fig:perf}. Like in section.~\ref{sec:data}, we could use standard histograms with bins of equal size. However, in this case the occupation number of the bins strongly depend on the assumed support of the distributions and on the binning. To avoid zero bins and sparse distributions we have to define the ranges and binnings by hand, making this strategy inconsistent in evaluation. Instead, we now split the support of the distributions into bins of equal probability weight, so-called quantiles, forming the set  $\sQ$. We generate the quantiles for a given distribution by iteratively dividing the set of validation showers into equal-sized subsets and keeping the median as the edge of the quantile. For multi-dimensional distributions, the splitting dimensions alternate. Figure~\ref{fig:quantiles} illustrates this algorithm. When comparing generated with reference samples, we want to increase the number of quantiles as far as possible, to cover the entire respective distribution at sufficient resolution.
    
In this iterative quantile scheme, zero bins will still occur once the number of quantiles exceeds the number of generated showers. To ensure the statistical fluctuations per bin are small and do not cause empty quantiles, we discard results for more than $n/10$ bins, where $n$ is the number of showers in the evaluation set.  This leads to roughly 10 events per bin, because the evaluated data is either generated from the same distribution as the validation data or is trained to resemble it well. As the event counts follow a Poisson distribution, the probability for a zero bin to occur can be calculated for the average occupation and gives around $4.5 \cdot 10^{-5}$.
    
\subsubsection*{Jensen--Shannon divergence}

The evaluation chain for the quality of the generated samples starts by constructing quantiles from the validation set. This defines our approximate truth density $p_i$ per quantile $i$. Next, we extract the density of showers $g_i$ per quantile, either for smaller sets of \geant showers or 1000k VAE-GAN generated showers. Due to values appearing in our validation set multiple times, quantiles are not uniquely defined, so the $p_i$ values may differ slightly from their constructed value $1/\#\sQ$.
    
To measure the similarity of the two distributions, we use the Jensen-Shannon divergence 
\begin{align}
        \jsd (g,p) = \frac{1}{2} \kl \left( g \,\Bigg | \, \frac{g+p}{2} \right) + \frac{1}{2} \kl \left( p \,\Bigg |  \, \frac{g+p}{2} \right) \; .
\end{align}
The $\jsd$ can be understood as a symmetrized version of the Kullback–Leibler (KL)-divergence
\begin{align}
    \kl(g \, | \,p) = \int g(x) \log \frac{g(x)}{p(x)} \, \mathrm{d}x.
\end{align}
For the VAE-GAN results, where $g = g(x)$ is the generated distribution, the $\jsd$ is the exact entity optimized by the $\min$-$\max$ training on the GAN loss defined in eq.~\eqref{eq:loss}~\cite{goodfellow, Buhmann:2020pmy}. For GAN and Monte Carlo methods, we usually do not have an explicit form of the generated distributions, but only sets $\sG$ and $\sP$ generated from the estimated distribution $g$ or the true distribution $p$. This is why we estimate the $\jsd$ for the continuous distributions from the quantile values
\begin{align}
        \bjsd (g,p) = \frac{1}{2} \sum_{Q_i \in \sQ} \left( g_i \log \frac{g_i}{\frac{1}{2}(g_i + p_i)} + p_i \log \frac{p_i}{\frac{1}{2}(g_i + p_i)}\right) \; .
\label{eq:bjsd}
\end{align}
Just like the $\jsd$, this estimate lies between zero and $\log2$. It turns into the continuous $\jsd$ between the histogram estimators 
\begin{align}
        \begin{aligned}
            \overline{g}(x) 
            &= \sum_{Q_i \in \sQ} \frac{g_i}{\mathrm{vol} (Q_i)} \, \mathrm{1}_{Q_i}(x) 
            = \sum_{Q_i \in \sQ} \frac{ \# \lbrace x' \in Q_i \,|\, x' \in \sG \rbrace}{\# \sG \cdot \mathrm{vol} (Q_i)} \, \mathrm{1}_{Q_i}(x) \\
            \text{and} \qquad 
            \overline{p}(x) 
            &= \sum_{Q_i \in \sQ} \frac{p_i}{\mathrm{vol} (Q_i)} \, \mathrm{1}_{Q_i}(x) \; ,
        \end{aligned}
    \end{align}
with $\mathrm{vol}$ the n-dimensional volume, $1_{Q_i}$ the indicator function of the i-th quantile and $\sG$ all showers in either an evaluation set of \geant samples or in the generated set. As for all histogram estimators, independent of the choice of bin edges, the overall number of bins, the cardinality of the fitted set, as well as the number of showers per bin have to go to infinity for the estimator to converge to the underlying distribution.  As $\bjsd$ goes to zero, the two distributions $g$ and $p$ are identical. \medskip

To determine the quality of our generative model relative to truth or validation distributions, we look at the dependence of the Jensen--Shannon divergence $\bjsd$ on the number of quantiles $\nq$ we can reliably construct. This will allow us to gauge where the density estimation underlying the VAE-GAN beats the statistically limited training data. As discussed earlier, we estimate the uncertainty on $\bjsd$ for the 5k and 10k evaluation sets of \geant data from five independent sets each.


\section{GANplification performance}
\label{sec:results}

Using our extended methodology we are now in a position to extend the toy study of ref.~\cite{Butter:2020qhk} to a relevant physics application, with the corresponding increased complexity and physics content.

\subsubsection*{Overcoming training statistics}

\begin{figure}[t]
    \centering
    \includegraphics[width=0.32\linewidth]{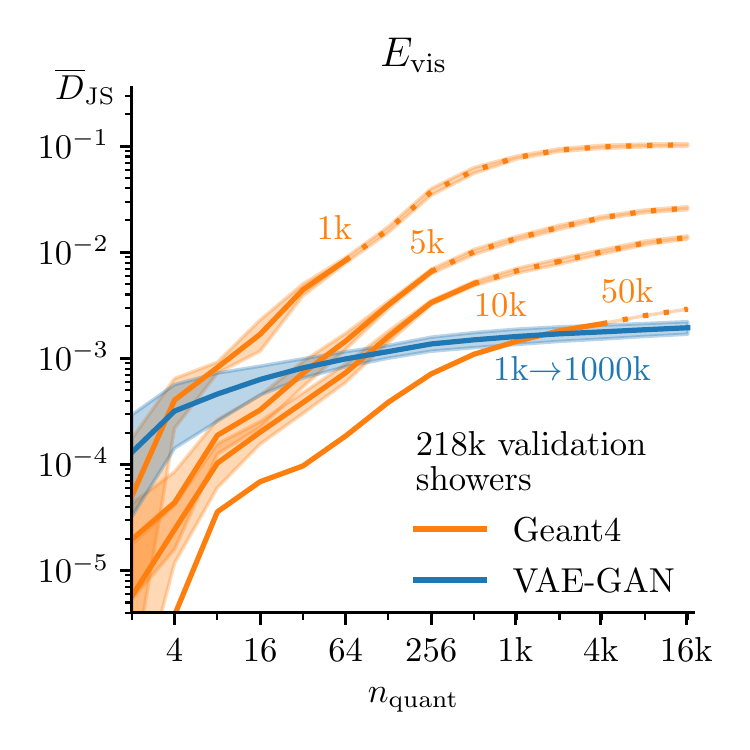}
    \includegraphics[width=0.32\linewidth]{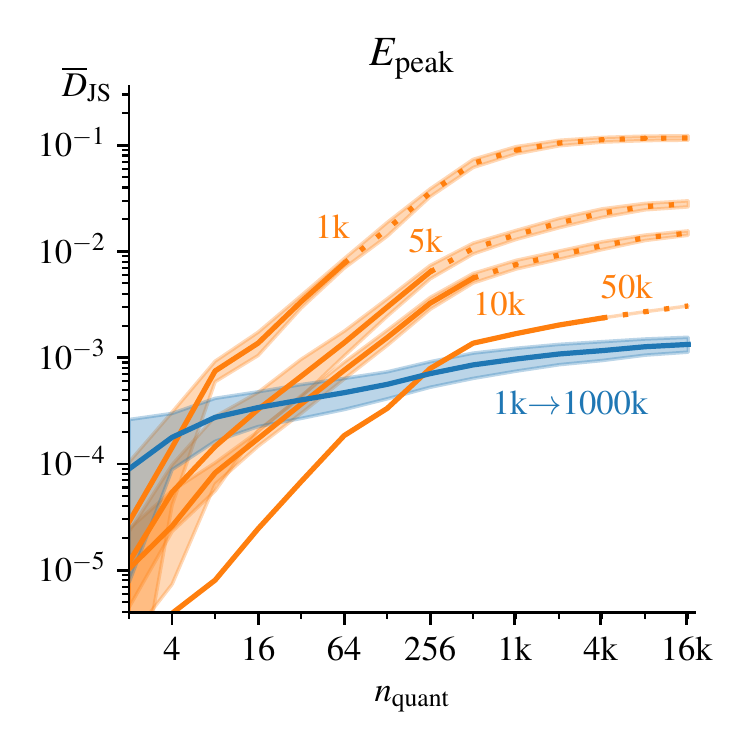}
    \includegraphics[width=0.32\linewidth]{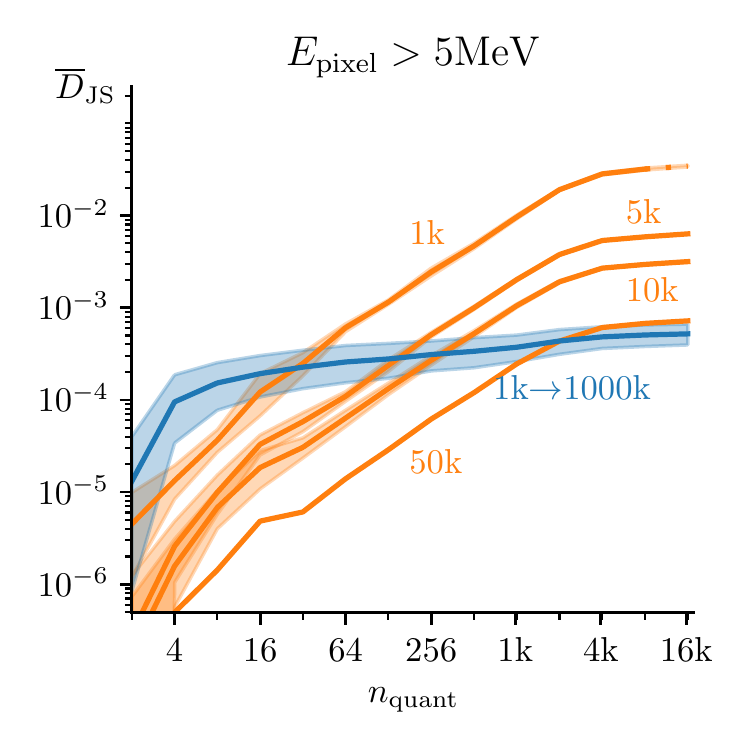}
    \includegraphics[width=0.32\linewidth]{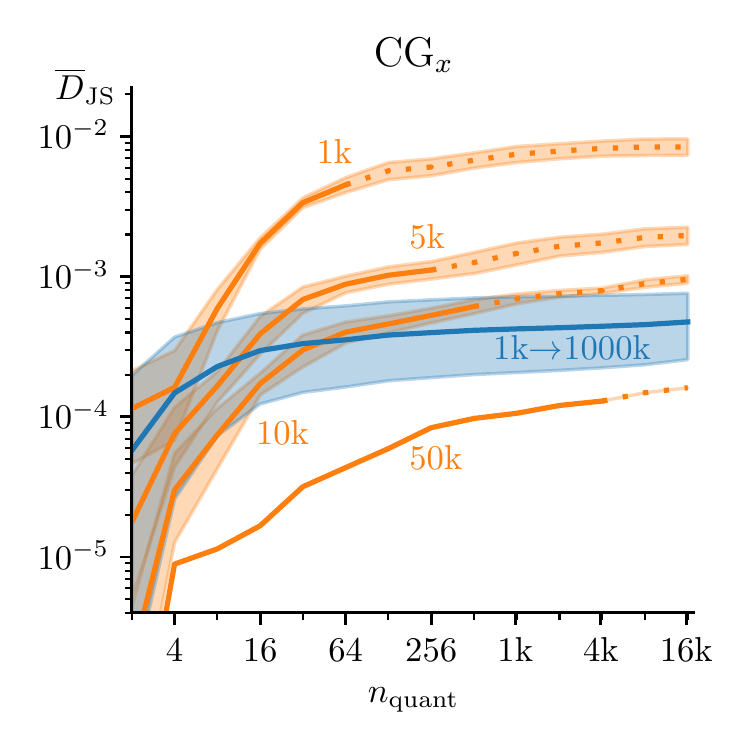}
    \includegraphics[width=0.32\linewidth]{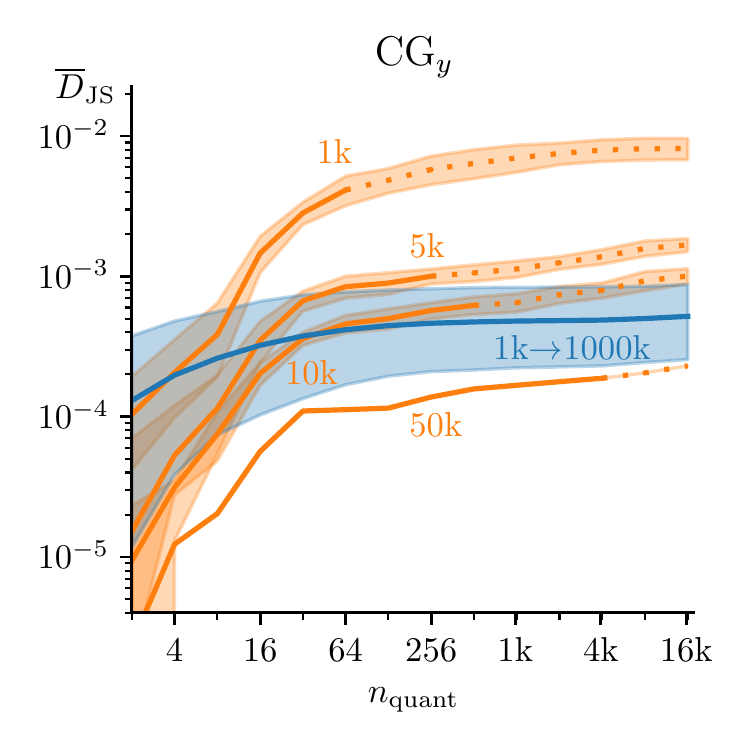}
    \caption{Dependence of $\bjsd$ on the number of quantiles $\nq$ for different amounts of \geant data (orange) and VAE-GAN data (blue) for the observables given in eq.~\eqref{eq:obs}. Solid lines indicate  meaningful, non-sparse quantile sets. The 1k \geant samples were also used to train the VAE-GAN. Errors are calculated as the standard deviation from five datasets. For 50k we omit the negligible errors.}
    \label{fig:comp_to_data_1d}
    \end{figure}

In figure~\ref{fig:comp_to_data_1d} we show how $\bjsd$ depends on the number of quantiles for the different observables given in eq.~\eqref{eq:obs}. For simple, uni-modal distributions like the energy sum, the peak energy and the centers of gravity, 1000k showers generated from the VAE-GAN achieve similar values as the 1k training data for very low numbers of bins. This means the generated data closely resembles the mean, standard deviation and low-level moments of the training data. For the more complex distribution of the visible per-pixel energy, the $\bjsd$ only resolves part of the high-density regions for a small number of quantiles. Increasing the numbers of quantiles, the interpolation of the generative model in the sparsely populated areas of the support starts to help, and the $\bjsd$-values for the \geant data increases over the VAE-GAN level. As there are on average about 13 active pixels above 5~MeV, as seen in figure~\ref{fig:perf}, the statistics for the per-pixel energy distribution benefits from these 13 pixel measurements per shower. For large numbers of quantiles, the $\bjsd$ values of the VAE-GAN are consistently below the corresponding values for the training sample and for all observables. This amplification is a result of the interpolation via the generative model's smoothing properties.
    
To quantify the amplification, we can compare the VAE-GAN distributions to larger \geant samples. Again, for small numbers of quantiles the VAE-GAN does not reach the truth $\bjsd$-values of larger data samples. This confirms that the neural network does not add global information to the training data and will not improve, for instance, the estimated mean of a Gaussian distribution. On the other hand, what we are really interested in are the features over the full distributions. In figure~\ref{fig:comp_to_data_1d} we show how the network trained on 1k showers and used to generate 1000k showers plateaus in $\bjsd$, as a function of the resolution, and how this plateau value compares to different \geant sample sizes. 
For a large number of quantiles and probing detailed features of the distributions, our VAE-GAN surrogate description corresponds to at least 50k \geant showers when we look at $E_\text{vis}$, $E_\text{peak}$, or $E_\text{pixel}$. This gives us GANplification factors as large as 50 for the relevant high-resolution features. For the reconstructed center of gravity this factor becomes a little smaller, but remains above ten.

\begin{figure}[t]
    \includegraphics[width=0.32\linewidth]{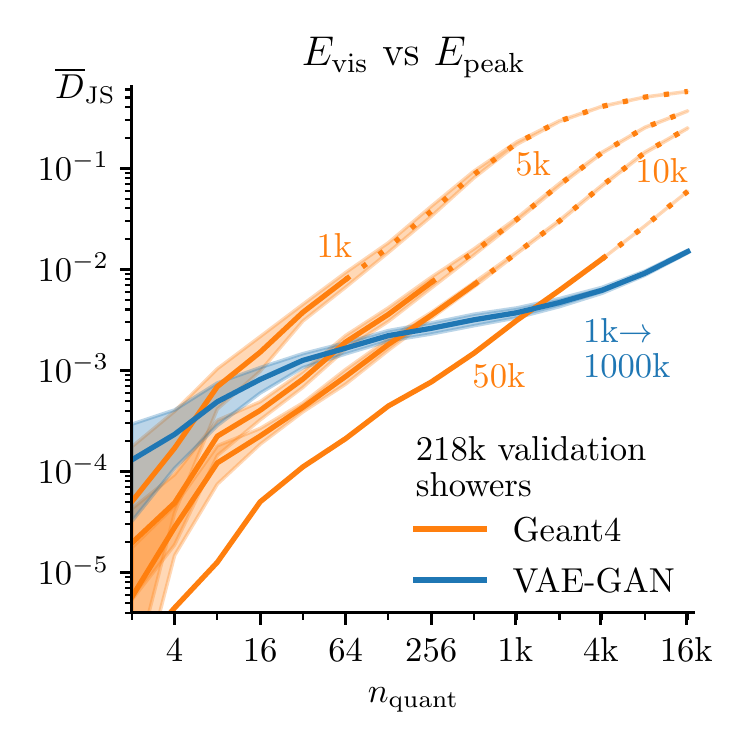}
    \includegraphics[width=0.32\linewidth]{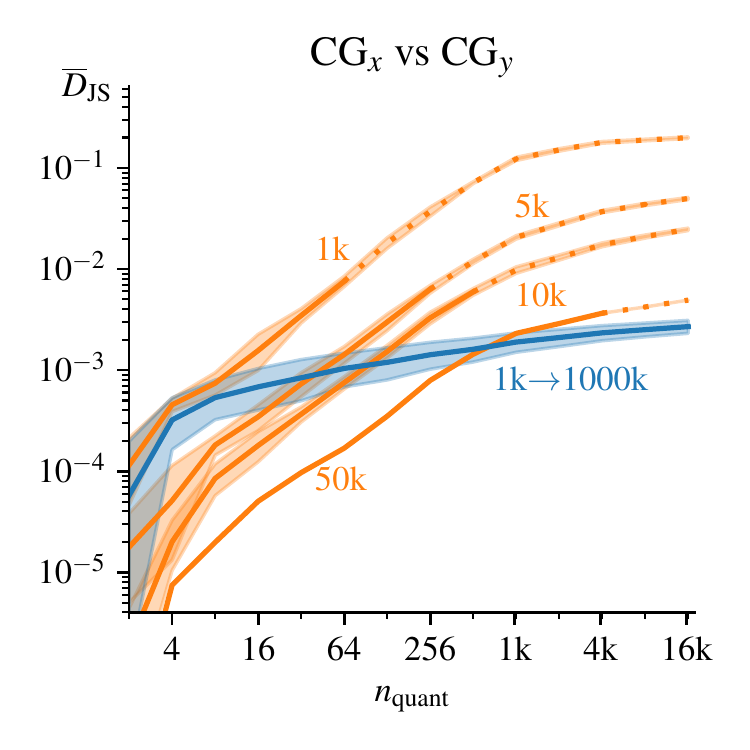}
    \includegraphics[width=0.32\linewidth]{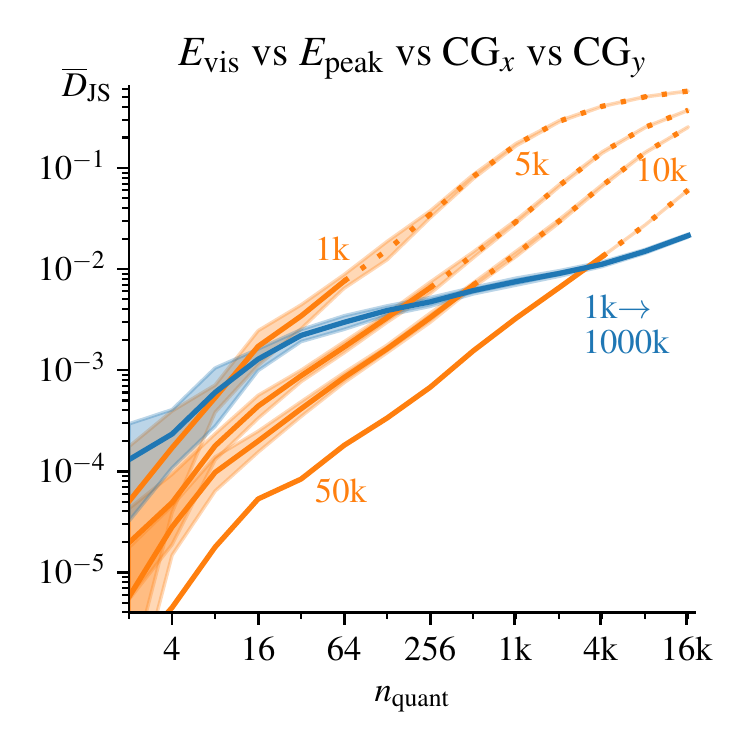}
    \caption{Dependence of $\bjsd$ on the number of quantiles $\nq$ for different amounts of \geant data (orange) and VAE-GAN data (blue), now for correlations between the observables of eq.~\eqref{eq:obs}, corresponding to the 1D results in figure~\ref{fig:comp_to_data_1d}.}
    \label{fig:comp_to_data_nd}
\end{figure}

Similar observations can be made for joint distributions, or correlations, of the different observables. Figure~\ref{fig:comp_to_data_nd} shows how the VAE-GAN encodes the correlations between observables with a consistently smaller error than the training data. The per-pixel energy distribution cannot be included in the correlations, as it features a varying number of pixel energy values per shower, whereas all other observables give a single value per shower. An unexpected upwards slope appears when examining joint distributions containing the energy sum and the peak energy of the generated images. This can be traced back to slight, small-scale fluctuations in the correlation between them in the generated data. Still, in all of the correlations we find a GANplification factor larger than 50 for the relevant detailed features, larger than for the one-dimensional distributions, as expected from the higher dimensionality and therefore reduced per-quantile statistics. 

\subsubsection*{Density estimation}

\begin{figure}[b]
    \includegraphics[width=0.95\linewidth]{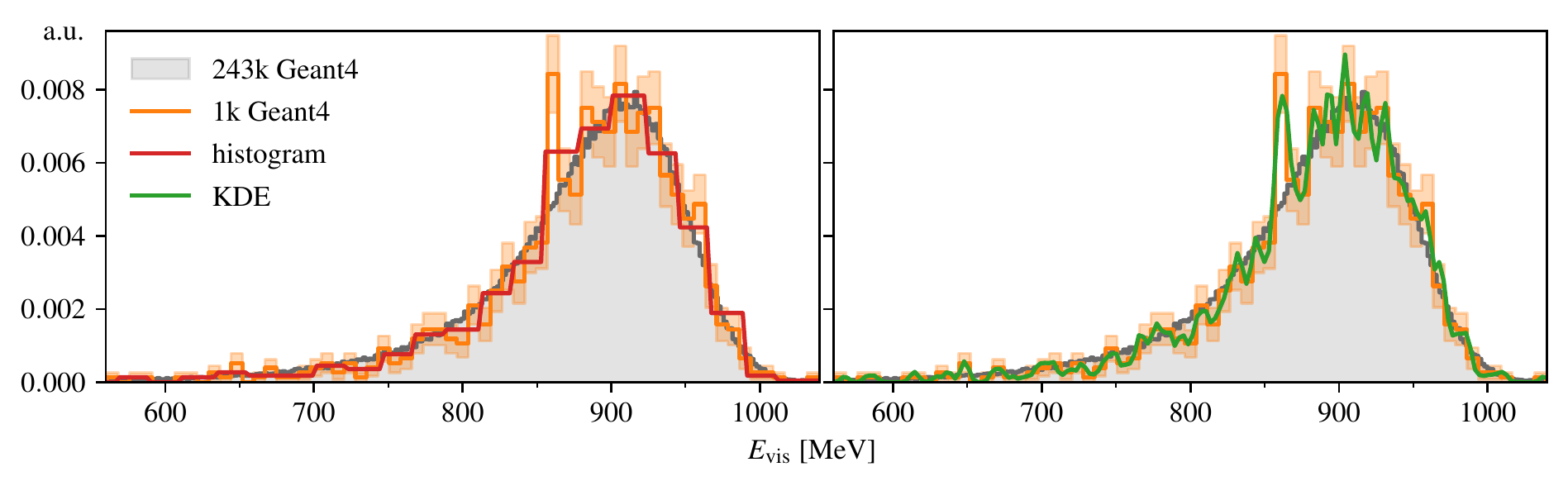}
    \caption{Example of an histogram estimator (red) and a kernel density estimator (green). The orange histogram shows the training data, including Poisson errors, that both estimators where fitted to using cross-validation.}
    \label{fig:est}
\end{figure}

\begin{figure}[t]
    \centering
    \includegraphics[width=0.33\linewidth]{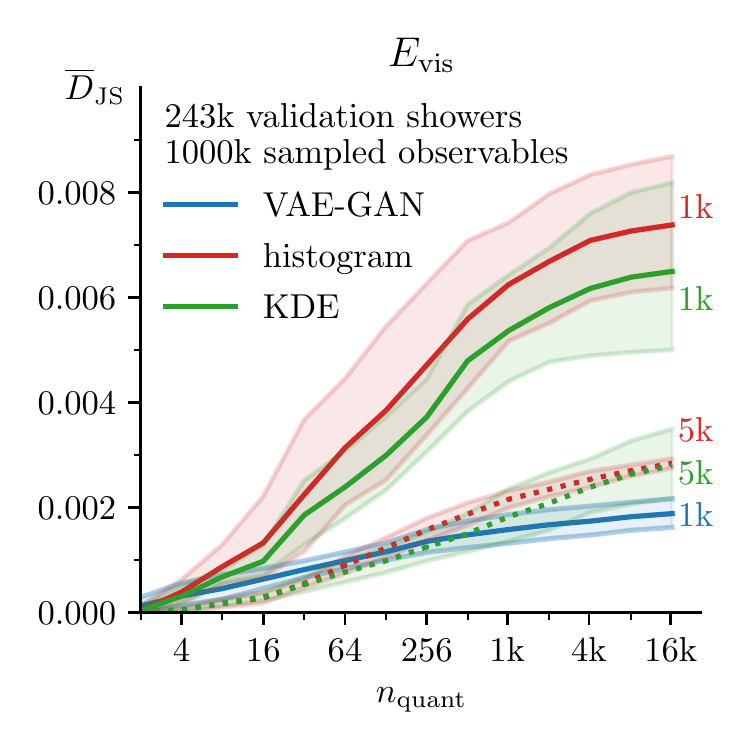}
    \includegraphics[width=0.33\linewidth]{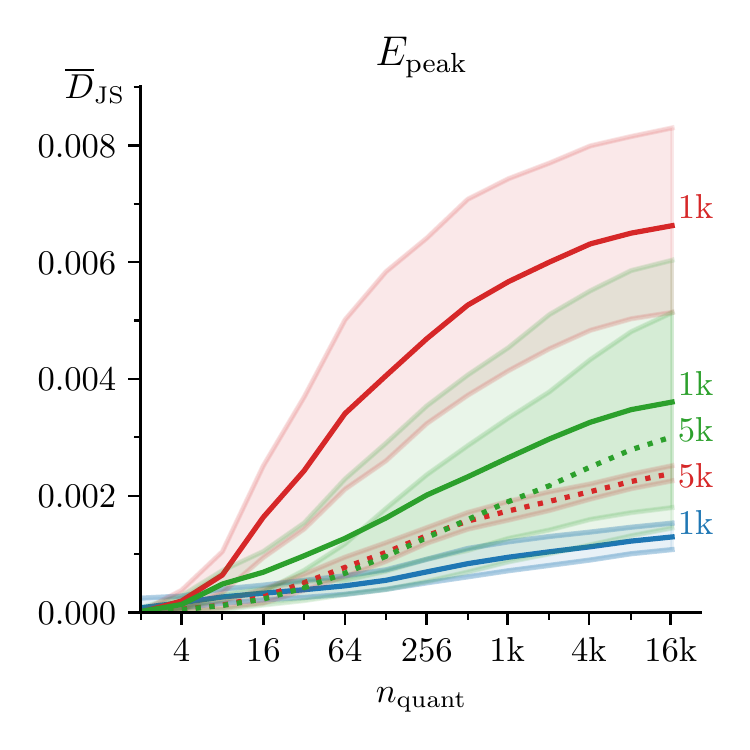} \\
    \includegraphics[width=0.33\linewidth]{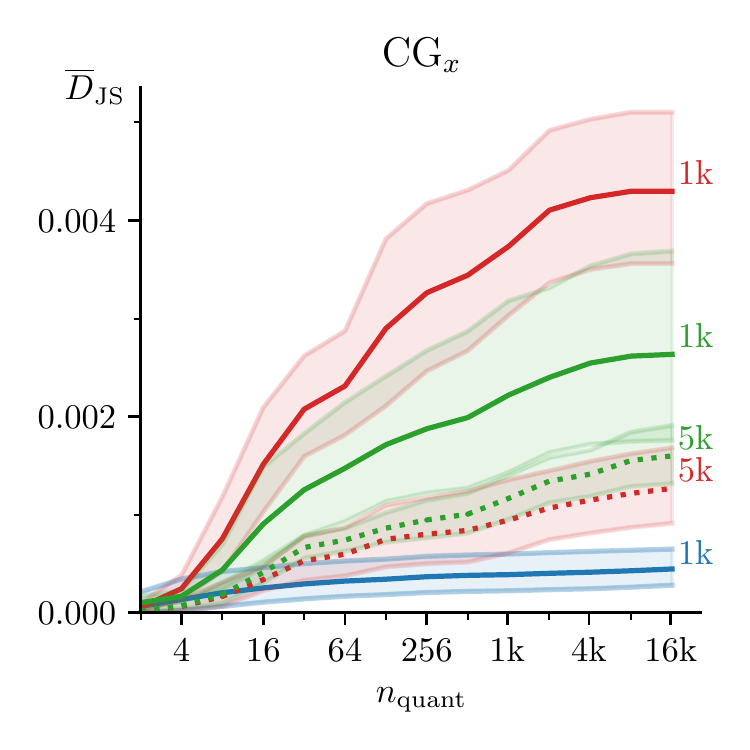}
    \includegraphics[width=0.33\linewidth]{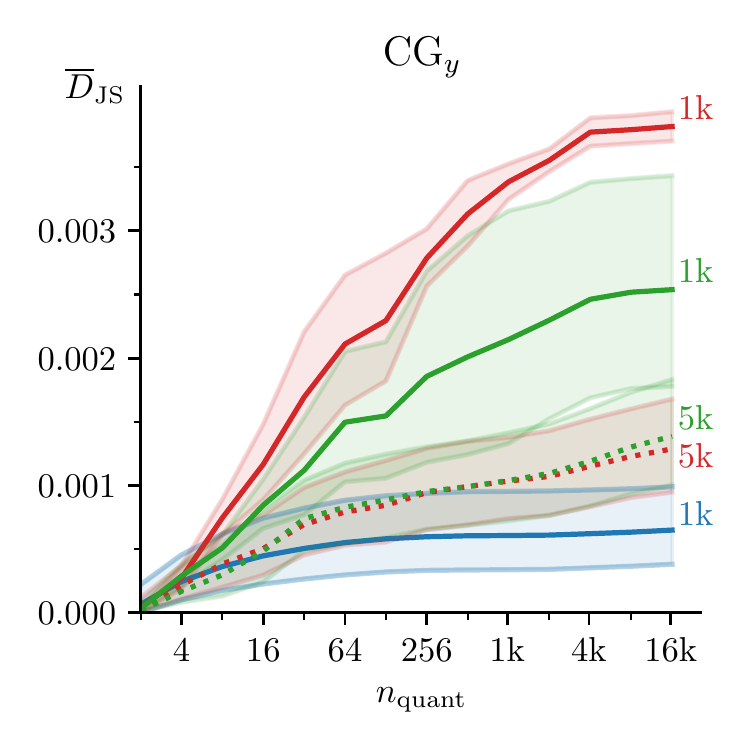}
    \caption{Dependence of $\bjsd$ on the number of quantiles $\nq$ for 1000k observable values sampled from histogram estimators (red) and kernel density estimators (green) and for 1000k showers sampled from VAE-GANs (blue). Errors are calculated as the standard deviation of five fits to different datasets. The size of the training sets is given to the right of the corresponding lines.}
    \label{fig:comp_to_est_1D}
\end{figure}

After we have seen that it is beneficial to generate datasets based on a learned density estimation, the question is whether other ways to estimate densities can give similar results. While there exists literature on convergence rates of generative methods~\cite{Biau2018, belomestny2021rates}, our physics application is defined by very specific limitations, different from those formal arguments. We therefore compare the performance of our VAE-GAN to two classical density estimation techniques. For both of them we analyze the same one-dimensional and multi-dimensional kinematic distributions as before. 

To each of our five training sets, we fit a kernel density estimator (KDE) and a histogram estimator, by minimizing the mean negative $\log$-likelihood of cross-validation subsets of the training set on a grid of the estimator parameters. The results for the energy sum are shown in figure~\ref{fig:est}. For the KDE we use the \sklearn~\cite{scikit-learn} \textsc{KernelDensity} class together with the built-in \textsc{GridSearchCV} tool using 5-fold cross-validation to optimize the bandwidth of the Gaussian kernel. The values of the bandwidth for the individual optimizations are given in table~\ref{tab:kde}. The parameters of the histogram estimator, \ie the number of bins along the individual dimensions, are optimized using our own implementation of the same techniques. To ensure stable convergence, we form 500 cross-validation sets from the training data. The results of this optimization can again be found in table~\ref{tab:kde}.
    
In figure~\ref{fig:est} we see that the KDE tends to over-fit and that the histogram estimator is  limited by its discrete functional form. We can analyze their performance more quantitatively using the $\bjsd$ shown in figure~\ref{fig:comp_to_est_1D}. Due to the logarithmic nature and complex functional form of the per-pixel energy distribution, the histogram estimator and the KDE do not converge for the low number of training showers we use, so we omit this observable. First, trained on 1k showers, the histogram estimator can only use very few bins to balance over-fitting against the approximation error caused by its coarse structure and is thus outperformed by the KDE. For a larger training set and the correspondingly larger number of bins, the approximation errors drop and both estimation methods perform similarly. However, compared to the VAE-GAN, both techniques lack descriptive power for small scales. Only for two to four bins they perform similarly to the VAE-GAN. Next, comparing the generative network to density estimators fitted to 5k showers, we can again observe the benefits of higher statistics for estimating low moments of the distributions. 

\begin{figure}[t]
    \includegraphics[width=0.32\linewidth]{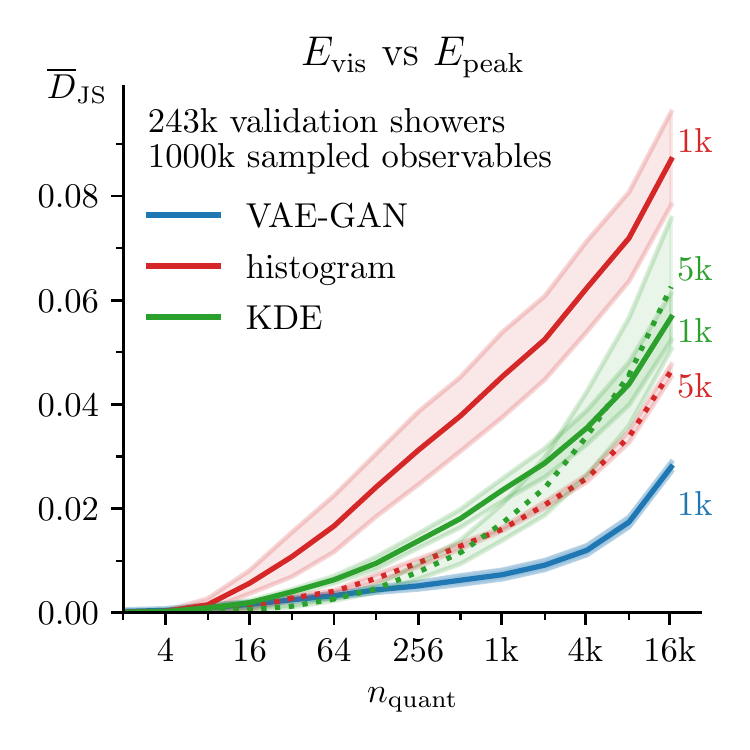}
    \includegraphics[width=0.32\linewidth]{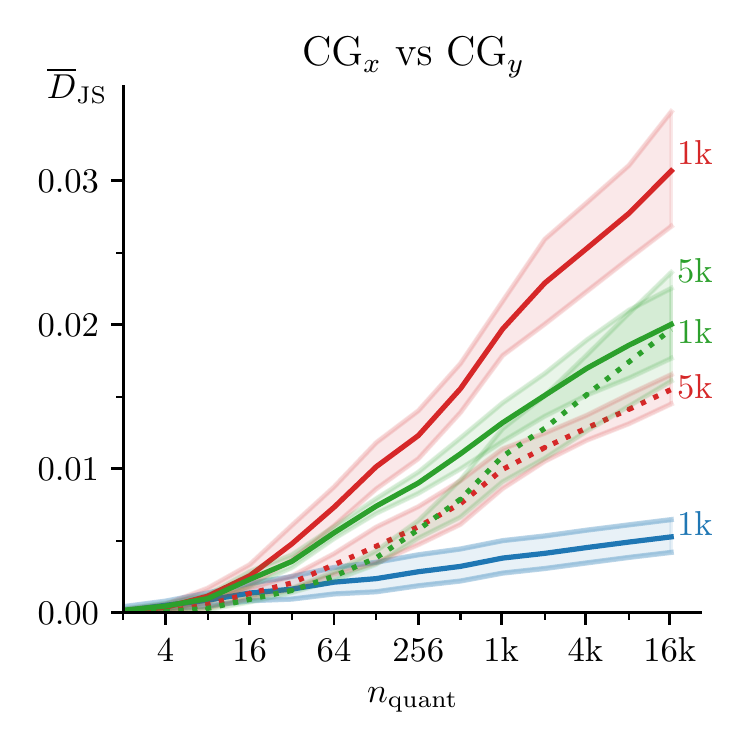}
    \includegraphics[width=0.32\linewidth]{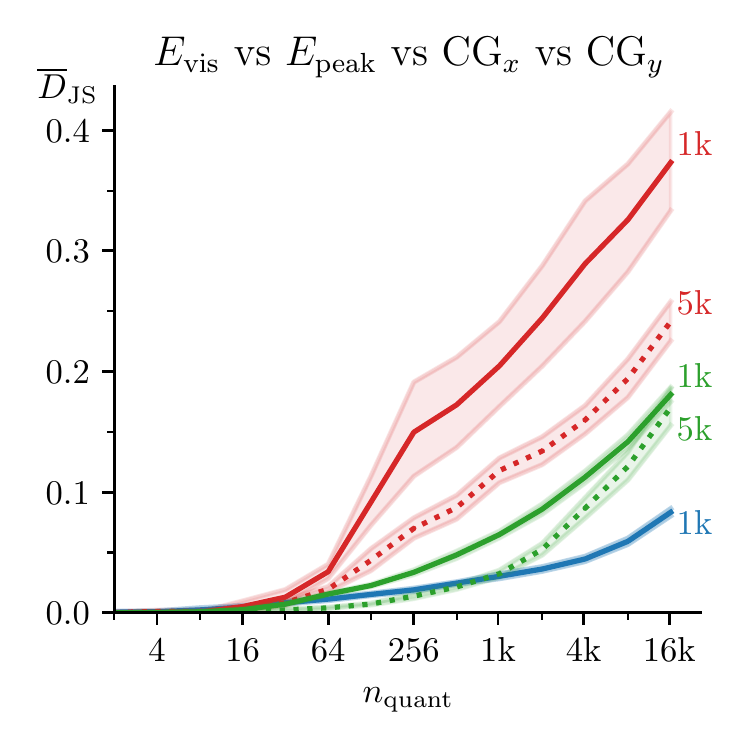}
    \caption{Dependence of $\bjsd$ on the number of quantiles $\nq$ for 1000k observable values sampled from different density estimators for multi-dimensional combinations of the observables given in eq.~\eqref{eq:obs}, in analogy to the 1D results in figure~\ref{fig:comp_to_est_1D}.}
    \label{fig:com_to_est_nD}
\end{figure}

For the 2-dimensional correlations shown in figure~\ref{fig:com_to_est_nD} we find similar limitations of the classical methods. Only the 4-dimensional density estimation behaves differently in that the histogram estimator is generally outperformed by the KDE. We can understand these patterns from the histogram parameters in table~\ref{tab:kde}. As the histogram estimator introduces bins in every direction, the number of showers per bin drops inversely proportional to the volume of the space. To avoid over-fitting, only few bins per dimension can then be used, leading to a large approximation error. The KDE and the VAE-GAN scale better with the number of dimensions, and as before the KDE only matches the VAE-GAN performance for a very small number of quantiles.

In addition to the neural network outperforming both density estimators, we remind ourselves that the VAE-GAN actually performs the more general task of estimating the distribution of calorimeter images or low-level observables, whereas the classical methods estimate the distributions of the high-level observables.

\begin{table}[H]
\centering
    \caption{KDE bandwidths and numbers of bins in the according dimensions for the histogram estimators presented in figures~\ref{fig:comp_to_est_1D} and~\ref{fig:com_to_est_nD}. Estimators are fitted for five independent training sets to extract the mean and standard deviation.}
    \label{tab:kde}
\vspace{0.5cm}
\begin{small} \begin{tabular}{l|cccc}
        \toprule
& KDE bandwidth 1k & KDE bandwidth 5k & \# histogram bins 1k & \# histogram bins 5k                     \\ \midrule
$E_\text{vis}$      & $0.05 \pm 0.01$  & $0.03 \pm 0.01$  & $39 \pm 10$ & $58 \pm 7$ \\ 
$E_\text{peak}$ & $0.10 \pm 0.03$  & $0.03 \pm 0.02$  & $27 \pm 4$  & $50 \pm 5$ \\
$\text{CG}_x$         & $0.10 \pm 0.02$  & $0.02 \pm 0.01$  & $32 \pm 12$ & $49 \pm 13$ \\
$\text{CG}_y$         & $0.10 \pm 0.02$  & $0.03 \pm 0.01$  & $25 \pm 4$  & $43 \pm 7$ \\ \midrule        
$E_\text{vis}$ vs $E_\text{peak}$ 
& $0.09 \pm 0.01$ & $0.03 \pm 0.01$  & $30 \pm 3 \times 26 \pm 7$ & $40 \pm 2 \times 46 \pm 5$ \\ 
$\text{CG}_x$ vs $\text{CG}_y$            
& $0.18 \pm 0.02$ & $0.07 \pm 0.01$  & $21 \pm 1 \times 20 \pm 1$ & $21 \pm 1 \times 22 \pm 2$ \\ 
complete 4D     
& $0.24\pm 0.01$  & $0.12 \pm 0.01$  & $20 \times 19 \times 5 \times 5 \pm 1$ & $21 \times 21 \times 7 \times 8 \pm 1$ \\ \bottomrule
\end{tabular} \end{small}
\end{table}

\section{Conclusions}
\label{sec:conclusions}

In this paper we have shown that a realistic generative ML-model can indeed be used to generate a large number of showers,  beyond a limited training statistics. Specifically, we used a VAE-GAN to generate photon showers for the electromagnetic calorimeter of the planned ILD detector design at a future linear collider. Our model is a simplification of the established precision-simulation network developed for this task~\cite{Buhmann:2020pmy}. This model is trained on a small number of showers from a \geant simulation, where a high-statistics sample of \geant showers serves as a truth estimate. Relative to this truth sample, we estimate the information content of finite-size samples using quantiles for standard kinematic observables and their correlations. A variable number of quantiles allows us to balance resolution with statistics.

Our study confirms earlier results based on a simple Gaussian example~\cite{Butter:2020qhk}, in that for a properly trained network a set of generated showers comparable in size to the training data provides a physics-wise nearly equivalent but statistically independent copy of the training data. More generated showers will, individually, contain less information than an actual shower, but add information as a sample. This amount of information can be linked to an effective sample size of actual data. For very large numbers of generated showers, the information in the generated sample reaches a plateau, reflecting limitations of the network architecture and training.

For our problem and network at hand, we find that the effective sample sizes give an enhancement or GANplification factor of 10 to 50, for a large number of quantiles and corresponding to high-resolution kinematic features. For a training sample of 1k showers we generate up to 1000k showers from the network and find a comparable performance of up to 50k \geant showers for the kinematic distributions and their correlations. We also interpret the VAE-GAN as a density estimator and find that it learns the truth density from the showers better than standard density estimators on the high-level kinematic variables. This proves that the generative network can even learn and sample from implicitly defined distributions and benefit from superior interpolation or fit properties. These properties motivate deep generative detector simulations for statistical amplification in addition to computational acceleration.

\acknowledgments

We would like to thank Suada Mulgeci for the valuable discussions in the earlier stages of this project.
The research of AB and TP is supported by the Deutsche Forschungsgemeinschaft (DFG, German Research Foundation) under grant 396021762 – TRR 257 Particle Physics Phenomenology after the Higgs Discovery. This work was supported by the Deutsche Forschungsgemeinschaft (DFG, German Research Foundation) under Germany's Excellence Strategy EXC 2181/1 - 390900948 (the Heidelberg STRUCTURES Excellence Cluster). BN is supported by the U.S. Department of Energy, Office of Science under contract DE-AC02- 05CH11231. SB is supported by the Helmholtz Information and Data Science Schools via DASHH (Data Science in Hamburg - HELMHOLTZ Graduate School for the Structure of Matter) with the grant HIDSS-0002. DH and SD acknowledge support by the Deutsche Forschungsgemeinschaft (DFG, German Research Foundation) under Germany’s Excellence Strategy – EXC 2121 „Quantum Universe“ – 390833306. EE is funded through the Helmholtz Innovation Pool project ACCLAIM that provided a stimulating scientific environment for parts of the research done here. This research was supported in part through the Maxwell computational resources operated at Deutsches Elektronen-Synchrotron DESY, Hamburg, Germany.


\bibliographystyle{JHEP}
\bibliography{literature}

\end{document}

%% file: settings_2.tex
\usepackage[utf8]{inputenc} 
\usepackage[english]{babel} 

\usepackage{geometry} 		
\usepackage{mathtools} 		
\usepackage{float} 			
\usepackage{tabularx} 		
\usepackage{booktabs} 		
\usepackage{color, xcolor} 	
\usepackage{pdfpages} 		
\usepackage{extarrows} 		
\usepackage{multirow} 		
\usepackage{multicol} 		
\usepackage{caption} 		
\usepackage{subcaption} 	
\usepackage{enumitem} 		
\usepackage{setspace} 		
\usepackage{xspace} 		
\usepackage{ragged2e} 		
\usepackage{stackrel} 		
\usepackage{tikz} 			
\usepackage{braket} 		
\usepackage{bm} 			
\usepackage{tensor} 		
\usepackage{slashed} 		
\usepackage{siunitx} 		
\usepackage{lastpage} 		
\usepackage[normalem]{ulem} 
\usepackage{fontawesome} 	
\usepackage{tocloft} 		
\usepackage{titlesec} 		
\usepackage[most]{tcolorbox} 					
\usepackage[nottoc, notlot, notlof]{tocbibind} 	
\usepackage[ruled, vlined]{algorithm2e} 		

\DeclareSymbolFont{usualmathcal}{OMS}{cmsy}{m}{n}
\DeclareSymbolFontAlphabet{\mathcal}{usualmathcal}

\usepackage{jinstpub}
\usepackage[nameinlink, capitalize]{cleveref} 	

%% file: shortcuts.tex
\newcommand{\ie}{\textsl{i.e.}\;}

\newcommand\one{\leavevmode\hbox{\small1\normalsize\kern-.33em1}}

\newcommand{\loss}{\mathcal{L}} 	

\newcommand{\sklearn}{\texttt{scikit-learn}\xspace}

\newcommand{\Adam}{\texttt{Adam}\xspace}

\newcommand{\geant}{\textsc{Geant4}\xspace}

\newcommand{\pytorch}{\textsc{PyTorch}\xspace}

\newcommand{\arXiv}[2][]{%
	\ifthenelse{\equal{#1}{}}%
	{\href{http://arxiv.org/abs/#2}{arXiv:#2}}%
	{\href{http://arxiv.org/abs/#2}{arXiv:#2~[#1]}}}

\def\slashchar#1{\setbox0=\hbox{$#1$}           
   \dimen0=\wd0                                 
   \setbox1=\hbox{/} \dimen1=\wd1               
   \ifdim\dimen0>\dimen1                        
      \rlap{\hbox to \dimen0{\hfil/\hfil}}      
      #1                                        
   \else                                        
      \rlap{\hbox to \dimen1{\hfil$#1$\hfil}}   
      /                                         
   \fi}

\newcommand{\jsd}{\ensuremath{D_\text{JS}}}
\newcommand{\bjsd}{\ensuremath{\overline{D}_\text{JS}}}
\newcommand{\kl}{\ensuremath{D_\text{KL}}}
\newcommand{\nq}{\ensuremath{n_\mathrm{quant}}}

\newcommand{\sQ}{\ensuremath{\mathbf{Q}}}
\newcommand{\sP}{\ensuremath{\mathbf{P}}}
\newcommand{\sG}{\ensuremath{\mathbf{G}}}

%% file: figures/lagan_gen.tex
\tikzstyle{vector} = [rectangle, minimum width=0.25cm, minimum height=2.3cm, text centered, draw=black, fill=orange!30]

\tikzset{
  pics/matrix/.style args={#1,#2,#3,#4}{
     code={
       \draw (0,0) -- (#4,0) -- (#4,#4) -- (0,#4) -- (0,0);
       \node[#3] (#1) at (1,1) {#2};
     }
  }
}

\tikzset{
  pics/picmatrix/.style args={#1,#2,#3,#4,#5}{
     code={
       \draw (0,0) -- (#4,0) -- (#4,#4) -- (0,#4) -- (0,0);
       \node[#3] (#1) at (0.5*#4,0.5*#4) {\includegraphics[width=#4cm]{#5}};
     }
  }
}

\tikzset{
  annotated cuboid/.pic={
    \tikzset{%
      every edge quotes/.append style={midway, auto},
      /cuboid/.cd,
      #1
    }
    \draw [every edge/.append style={pic actions, densely dashed, opacity=.5}, pic actions]
    (0,0,0) coordinate (o) -- ++(-\cubescale*\cubex,0,0) coordinate (a) -- ++(0,-\cubescale*\cubey,0) coordinate (b) edge coordinate [pos=1] (g) ++(0,0,-\cubescale*\cubez)  -- ++(\cubescale*\cubex,0,0) coordinate (c) -- cycle
    (o) -- ++(0,0,-\cubescale*\cubez) coordinate (d) -- ++(0,-\cubescale*\cubey,0) coordinate (e) edge (g) -- (c) -- cycle
    (o) -- (a) -- ++(0,0,-\cubescale*\cubez) coordinate (f) edge (g) -- (d) -- cycle;
  },
  /cuboid/.search also={/tikz},
  /cuboid/.cd,
  width/.store in=\cubex,
  height/.store in=\cubey,
  depth/.store in=\cubez,
  units/.store in=\cubeunits,
  scale/.store in=\cubescale,
  width=10,
  height=10,
  depth=10,
  units=,
  scale=.1,
}

\tikzstyle{arrow} = [thick,shorten >=40pt,->,>=stealth]

\begin{tikzpicture}
    \node (z) [vector, yshift=-.2cm] {};
    \node [text width=2.2cm, below of=z, yshift=-.5cm] {$z\sim\mathcal{N}(0,\,1)$};
    
    \def\imf{1.5}
    \def\chf{0.16}
    \def\chftwo{0.4}
    \pic at (3.2,0) {annotated cuboid={width=128*\chf, height=3*\imf, depth=3*\imf}};
    \pic at (5,0) {annotated cuboid={width=64*\chf, height=3*\imf, depth=3*\imf}};
    \pic at (7,0) {annotated cuboid={width=64*\chf, height=6*\imf, depth=6*\imf}};
    \pic at (8.5,0) {annotated cuboid={width=6*\chftwo, height=6*\imf, depth=6*\imf}};
    \pic at (10,0) {annotated cuboid={width=6*\chftwo, height=12*\imf, depth=12*\imf}};
    \pic at (12,0) {annotated cuboid={width=6*\chftwo, height=10*\imf, depth=10*\imf}};
    \def\xpos{14}
    \def\ypos{-1.3}
    \def\groesse{1.7}
    \draw (\xpos+.4,\ypos+.4) pic{picmatrix={linear ,  , white, \groesse, figures/geant10x10.png}};
    \draw (\xpos+.3,\ypos+.3) pic{picmatrix={linear ,  , white, \groesse, figures/geant10x10.png}};
    \draw (\xpos+.2,\ypos+.2) pic{picmatrix={linear ,  , white, \groesse, figures/geant10x10.png}};
    \draw (\xpos+.1,\ypos+.1) pic{picmatrix={linear ,  , white, \groesse, figures/geant10x10.png}};
    \draw (\xpos,\ypos) pic{picmatrix={linear ,  , white, \groesse, figures/geant10x10.png}};
    \node at (19,0) [text width=2cm] { };
    
    \node at (.4,1.3) [text width=2cm] {1x1x30};
    \node at (2.8,.5) [text width=2cm] {128x3x3};
    \node at (5.1,.5) [text width=2cm] {64x3x3};
    \node at (7.2,.6) [text width=2cm] {64x6x6};
    \node at (9.2,.6) [text width=2cm] {6x6x6};
    \node at (10.7,.89) [text width=2cm] {6x12x12};
    \node at (12.7,.82) [text width=2cm] {6x10x10};
    \node at (15.6,1.1) [text width=2cm] {1x10x10};
    
    \node at (1.,-2.5) [text width=2.3cm, align=center] {\textbf{Linear Layer \\ Reshaping}};
    \node at (3.7,-2.5) [text width=2.3cm, align=center] {\textbf{Convolution \\ Leaky ReLU \\ Batch Norm}};
    \node at (5.4,-1.5) [text width=2cm, align=center] {\textbf{Upsample}};
    \node at (7.4,-2.5) [text width=3cm, align=center] {\textbf{LC Layer \\ Leaky ReLU \\ Batch Norm}};
    \node at (8.7,-1.5) [text width=2cm, align=center] {\textbf{Upsample}};
    \node at (10.85,-2.5) [text width=2.3cm, align=center] {\textbf{LC Layer \\Leaky ReLU}};
    \node at (13.3,-2.5) [text width=2cm, align=center] {\textbf{LC Layer \\ ReLU}};
    
\end{tikzpicture}

%% file: figures/lagan_disc.tex
\tikzstyle{vector} = [rectangle, minimum width=0.25cm, minimum height=2.3cm, text centered, draw=black, fill=orange!30]

\tikzset{
  pics/matrix/.style args={#1,#2,#3,#4}{
     code={
       \draw (0,0) -- (#4,0) -- (#4,#4) -- (0,#4) -- (0,0);
       \node[#3] (#1) at (1,1) {#2};
     }
  }
}

\tikzset{
  pics/picmatrix/.style args={#1,#2,#3,#4,#5}{
     code={
       \draw (0,0) -- (#4,0) -- (#4,#4) -- (0,#4) -- (0,0);
       \node[#3] (#1) at (0.5*#4,0.5*#4) {\includegraphics[width=#4cm]{#5}};
     }
  }
}

\tikzset{
  annotated cuboid/.pic={
    \tikzset{%
      every edge quotes/.append style={midway, auto},
      /cuboid/.cd,
      #1
    }
    \draw [every edge/.append style={pic actions, densely dashed, opacity=.5}, pic actions]
    (0,0,0) coordinate (o) -- ++(-\cubescale*\cubex,0,0) coordinate (a) -- ++(0,-\cubescale*\cubey,0) coordinate (b) edge coordinate [pos=1] (g) ++(0,0,-\cubescale*\cubez)  -- ++(\cubescale*\cubex,0,0) coordinate (c) -- cycle
    (o) -- ++(0,0,-\cubescale*\cubez) coordinate (d) -- ++(0,-\cubescale*\cubey,0) coordinate (e) edge (g) -- (c) -- cycle
    (o) -- (a) -- ++(0,0,-\cubescale*\cubez) coordinate (f) edge (g) -- (d) -- cycle;
  },
  /cuboid/.search also={/tikz},
  /cuboid/.cd,
  width/.store in=\cubex,
  height/.store in=\cubey,
  depth/.store in=\cubez,
  units/.store in=\cubeunits,
  scale/.store in=\cubescale,
  width=10,
  height=10,
  depth=10,
  units=,
  scale=.1,
}

\tikzstyle{arrow} = [thick,shorten >=40pt,->,>=stealth]

\begin{tikzpicture}
    \def\xpos{-1.0}
    \def\ypos{-1.5}
    \def\groesse{1.7}
    \draw (\xpos+.4,\ypos+.4) pic{picmatrix={linear ,  , white, \groesse, figures/geant10x10.png}};
    \draw (\xpos+.3,\ypos+.3) pic{picmatrix={linear ,  , white, \groesse, figures/geant10x10.png}};
    \draw (\xpos+.2,\ypos+.2) pic{picmatrix={linear ,  , white, \groesse, figures/geant10x10.png}};
    \draw (\xpos+.1,\ypos+.1) pic{picmatrix={linear ,  , white, \groesse, figures/geant10x10.png}};
    \draw (\xpos,\ypos) pic{picmatrix={linear ,  , white, \groesse, figures/geant10x10.png}};
    
    \def\imf{1.7}
    \def\chf{0.22}
    \def\chftwo{0.35}
    \pic at (3.5,0) {annotated cuboid={width=32*\chf, height=10*\imf, depth=10*\imf}};
    \pic at (5.5,0) {annotated cuboid={width=8*\chftwo, height=8*\imf, depth=8*\imf}};
    \pic at (7.5,0) {annotated cuboid={width=8*\chftwo, height=6*\imf, depth=6*\imf}};
    \pic at (9.5,0) {annotated cuboid={width=8*\chftwo, height=6*\imf, depth=6*\imf}};
    \pic at (11.5,0) {annotated cuboid={width=8*\chftwo, height=3*\imf, depth=3*\imf}};
    \def\xpos{13.3}
    \def\ypos{-.5}
    \def\groesse{.3}
    \draw (\xpos,\ypos) pic{matrix={linear ,  , white, \groesse}};
    \draw (\xpos+.05,\ypos+.05) pic{matrix={linear ,  , white, \groesse}};
    \draw (\xpos+.1,\ypos+.1) pic{matrix={linear ,  , white, \groesse}};
    \draw (\xpos+.15,\ypos+.15) pic{matrix={linear ,  , white, \groesse}};
    \draw (\xpos+.2,\ypos+.2) pic{matrix={linear ,  , white, \groesse}};
    \draw (\xpos+.25,\ypos+.25) pic{matrix={linear ,  , white, \groesse}};
    \node at (17,0) [text width=2cm] { };
    
    \node at (.6,1.) [text width=2cm] {1x10x10};
    \node at (4,1.) [text width=2cm] {32x10x10};
    \node at (6.4,.9) [text width=2cm] {8x8x8};
    \node at (8.1,.75) [text width=2cm] {8x6x6};
    \node at (10.2,.75) [text width=2cm] {8x6x6};
    \node at (12.2,.65) [text width=2cm] {8x3x3};
    \node at (14.15,.5) [text width=2cm] {1x1x1};
    
    \node at (1.6,-2.6) [text width=2.3cm, align=center] {\textbf{Convolution \\ Leaky ReLU}};
    \node at (6.4,-2.5) [text width=2.6cm, align=center] {\textbf{3x \\ LC Layer \\ Leaky ReLU \\ Spectral Norm}};
    \node at (10.6,-1.4) [text width=2cm, align=center] {\textbf{Average Pooling}};
    \node at (12.9,-2.7) [text width=5cm, align=center] {\textbf{Minibatch Discrimination \\ Concat Diff \\ Linear Layer}};
\end{tikzpicture}